\documentclass[twocolumn]{aastex63}

\usepackage[caption=false]{subfig}
\usepackage{fontawesome}
\usepackage{CJK}

\usepackage{graphicx}
 \usepackage{amsmath}
\begin{document}
\begin{CJK*}{UTF8}{gbsn}
\title{The First Evidence of a Host Star Metallicity Cut-off In The Formation of Super-Earth Planets}

\email{boley.62@osu.edu}

\author[0000-0001-8153-639X]{Kiersten M. Boley}
\altaffiliation{NSF Graduate Research Fellow}
\affiliation{Department of Astronomy, The Ohio State University, Columbus, OH 43210, USA}

\author[0000-0002-8035-4778]{Jessie L. Christiansen}
\affiliation{Caltech/IPAC-NASA Exoplanet Science Institute, Pasadena, CA 91125, USA}

\author[0000-0003-1848-2063]{Jon Zink}
\affiliation{Department of Astronomy, California Institute of Technology, Pasadena, CA 91125, USA}
\altaffiliation{NHFP Sagan Fellow}

\author[0000-0003-3702-0382]{Kevin Hardegree-Ullman}
\affiliation{Department of Astronomy, The University of Arizona, Tucson, AZ 85721, USA}

\author[0000-0002-1228-9820]{Eve J. Lee}
\affiliation{Department of Physics and Trottier Space Institute, McGill University, Montreal, QC, H3A 2T8, Canada}

\author[0000-0003-3729-1684]{Philip F. Hopkins}
\affiliation{TAPIR, MS 350-17, Caltech, 1200 E. California Blvd, Pasadena, CA 91125}

\author[0000-0002-4361-8885]{Ji Wang (王吉)}
\affiliation{Department of Astronomy, The Ohio State University, Columbus, OH 43210, USA}

\author[0000-0002-3853-7327]{Rachel B. Fernandes}
\altaffiliation{President's Postdoctoral Fellow}
\affil{Department of Astronomy \& Astrophysics, 525 Davey Laboratory, The Pennsylvania State University, University Park, PA 16802, USA}
\affil{Center for Exoplanets and Habitable Worlds, 525 Davey Laboratory, The Pennsylvania State University, University Park, PA 16802, USA}

\author[0000-0003-4500-8850]{Galen J. Bergsten}
\affil{Lunar and Planetary Laboratory, The University of Arizona, Tucson, AZ 85721, USA}

\author[0000-0002-6673-8206]{Sakhee Bhure}
\affil{Centre for Astrophysics, University of Southern Queensland, Toowoomba, QLD 4350, Australia}




\begin{abstract}
Planet formation is expected to be severely limited in disks of low metallicity, owing to both the small solid mass reservoir and the low opacity accelerating the disk gas dissipation. While previous studies have found a weak correlation between the occurrence rates of small planets ($\lesssim$4$R_\oplus$) and stellar metallicity, so far no studies have probed below the metallicity limit beyond which planet formation is predicted to be suppressed. Here, we constructed a large catalog of $\sim$110,000 metal-poor stars observed by the TESS mission with spectroscopically-derived metallicities, and systematically probed planet formation within the metal-poor regime ([Fe/H] $\leq-0.5$) for the first time. Extrapolating known higher-metallicity trends for small, short-period planets predicts the discovery of $\sim$68 super-Earths around these stars { ($\sim$ 85,000 stars)} after accounting for survey completeness; however, we detect none. As a result, we have placed the most stringent upper limit on super-Earth occurrence rates around metal-poor stars {(-0.75 $<$ [Fe/H] $\leq$ -0.5)} to date, $\leq$ 1.67\%, a statistically significant (p-value = 0.000685) deviation from the prediction of metallicity trends derived with \emph{Kepler} and \emph{K2}. We find a clear host star metallicity cliff for super-Earths that could indicate the threshold below which planets are unable to grow beyond an Earth-mass at short orbital periods. This finding provides a crucial input to planet formation theories, and has implications for the small planet inventory of the Galaxy and the galactic epoch at which the formation of small planets started.

\end{abstract}

\section{Introduction} 

Under the core accretion paradigm, planet formation begins from the coagulation of solid material. Once this rocky core becomes massive enough (i.e., when its Bondi radius exceeds the core radius), gas accretion begins \citep[e.g.,][]{Pollack1996}, and the mass of the core ultimately determines the amount of accreted gas \citep[e.g.,][]{Lee2014,Lee2015}. This leads to the expectation of a strong correlation between the occurrence rate of gas giants and the amount of solid material in the disk, for which the metallicity of the host star is an excellent proxy \citep{Johnson2012,Hasegawa2014,Lee2019}. This giant planet-metallicity correlation has been well constrained from radial velocity surveys \citep{gon1997,Fischer2005,Udry2007,Johnson2010, Wang2015}. On the other hand, the super-Earth-metallicity correlation is expected to be weaker \citep{Lee2019}, and indeed only emerges in the much larger sample available via the \emph{Kepler} transit survey \citep{Kutra2021,Lu2020,Zhu2019,Petigura2018, Zink2023}.

While the observed correlation is weak, there are many theoretical reasons to suggest that there may be a critical threshold metallicity below which the formation of super-Earths becomes difficult, if not impossible. A lower limit to metallicity could arise from the physics of dust grain-grain collisions and dust growth coupled with metallicity-dependent disk evolution \citep{Johnson2012}, instabilities believed to trigger planetesimal formation \citep{Youdin2005,Johansen2007,Bai2010,Squire2018,Li2021}, or the necessity for planetary protocores to accrete solids from their environment via planetesimal \citep{Kokubo1998,Goldreich2004} or pebble accretion \citep{Ormel2010,Lambrechts2014,Lin2018}.
While the qualitative expectation of \emph{some} threshold is generic to most models, actual predictions of where that threshold should lie vary significantly with values from  
0.003 times solar metallicity \citep[see][their equation 10, evaluated at 0.1 AU]{Johnson2012} to solar or supersolar metallicities \citep[see][their equation 14]{Li2021} although the latter value is sensitive to the Stokes number of the coagulating dust grains (e.g., the critical minimum metallicity can be lower than the solar value if the Stokes number is large $\sim$0.1).

Because of the well-studied stellar age--metallicity correlation in the Galaxy, the value of this critical metallicity threshold directly impacts when the earliest small planets were formed in the Galaxy $-$ the oldest stars have metallicities significantly below the predicted range of values. Therefore, they would not be expected to form these planets. The super-Earth-metallicity correlation measured to date has been limited to host stars with metallicities greater than $[\mathrm{Fe/H}]=-0.4$, due to the focus of \emph{Kepler} on Sun-like stars, and the relative rarity of metal-poor stars in the local solar neighborhood. 

To date, there has been no large, systematic search for small planets orbiting metal-poor host stars. However, the Transiting Exoplanet Survey Satellite \citep[TESS,][]{TESS} enables consideration of the long-standing question: what is the critical metallicity required for small planet core formation? Given that TESS is an all-sky survey providing high-precision photometry on millions of stars, the number of metal-poor stars surveyed is orders of magnitude greater than its predecessors, \emph{K2} and \emph{Kepler}, enabling large population studies. Here, we perform the first analysis of the occurrence rates of small planets at metallicities approaching the predicted cut-offs for planet formation. We empirically test whether a metallicity cut-off exists and quantitatively address the metallicity below which such a drop in occurrence rate arises.

The outline of our paper is as follows: Section \ref{s:stellar} provides a description of our stellar sample. In Section \ref{s:detection}, we discuss our planet detection pipeline.  Section \ref{s:forward} contains a description of our forward model software. We discuss our results in Section \ref{s:results}. In Section \ref{s:discuss}, we compare our results to previous studies and their implications for super-Earth formation before summarizing our conclusions in Section \ref{s:conclusions}

\begin{figure}
\begin{center}
\includegraphics[width=\linewidth]{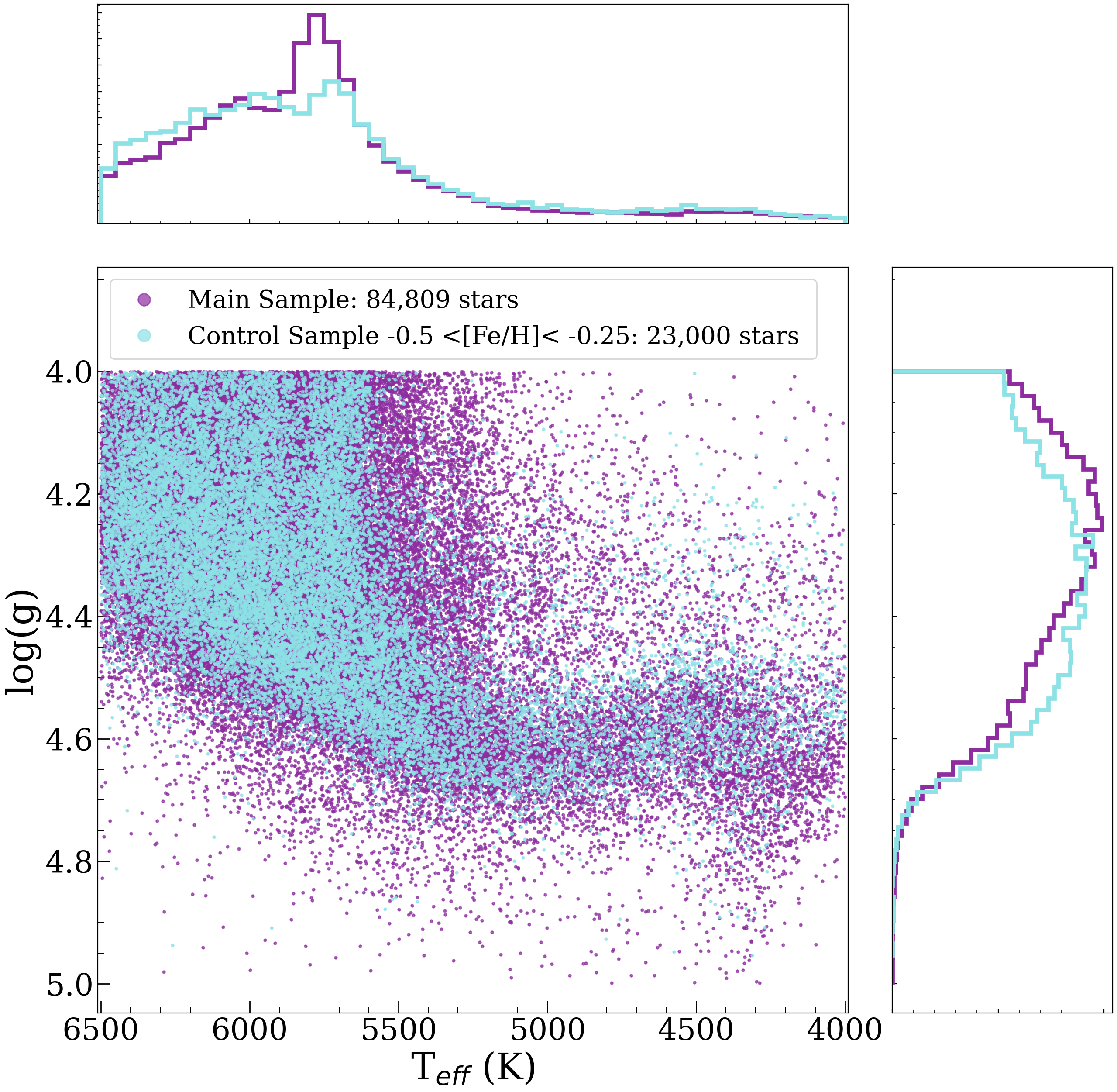}
\caption{The stellar sample from TESS portrayed in the $T_{eff}$ and log(g) plane. We show the main sample (-1$\leq$[Fe/H] $\leq$ -0.5) in purple, and the sub-sample (-0.5$<$[Fe/H] $\leq$ -0.25) in blue.\label{fig:stellar}}
\end{center}
\end{figure}

\section{Stellar Sample}\label{s:stellar}
Using the TESS Input Catalog ~\citep[TIC;][]{TESS}, we constructed a sample of metal-poor stars. We included TESS data from Sectors 1--55, and apply the following cuts: 

\begin{enumerate}
    \item \textbf{Spectroscopically-derived metallicities}: The TIC compiled data from many surveys into a catalog of $\sim$1.8 billion sources; thus, the spectroscopic metallicities are heterogeneous. The vast majority of these sources have parameters derived solely from photometric surveys, but the TIC also compiled data from the available spectroscopic surveys. To test existing planet formation theories, we included all stars with spectroscopically derived parameters having $[\mathrm{Fe/H}]<-0.5$, hereafter known as the main sample. In addition, we included a subset of stars with metallicities between $-0.5$ and $-0.25$ to compare and cross-check our TESS planet occurrence rates with previous results from \emph{Kepler} and \emph{K2}, referred to as the control sample hereafter. From this initial cut, we began with a main sample of 654,675 stars and control sample of 987,924 stars.
    \item \textbf{Stellar effective temperature}: We limited our sample to stars with effective temperatures ranging from 4000--6500~K. This requirement selects FGK spectral types and allows for direct comparison with previous higher metallicity \emph{Kepler} and \emph{K2} results. From this requirement, we excluded 32,609 main sample stars and 93,694 control sample stars. 
    \item \textbf{Surface gravity}: We excluded low surface gravity red giants by requiring $\log g > 4.0$ and removed 291,095 main sample stars and 377,648 control sample stars.
    \item \textbf{TESS magnitude ($T_{\rm mag}$)}: We selected stars brighter than $T_{\rm mag}=14$; at fainter magnitudes, our detection efficiency for small planets declines rapidly. From this cut, we excluded 232,312 main sample stars and 271,130 control sample stars.
    \item \textbf{Galactic Plane}: We excluded stars with $|b| <5^\circ$ to mitigate the effect of false positives due to crowding near the Galactic plane, removing 2,265 main sample stars and 33,869 control sample stars. 
    \item \textbf{Removal of Binary Stars}: Similar to previous works, we relied on the Gaia Renormalized Unit Weight Error (RUWE) metric to minimize this
    potential source of contamination and only included targets with RUWE$<$1.4 \citep{Stassun2021,Lindegren2018}.
    Additionally, Gaia DR3 provides a flag, ``non\_single\_star'', which denotes sources that provide evidence of a binary \citep{GaiaDR3}.
    Removing targets with Gaia binary flags and high RUWE values, we excluded 20,867 potential binary targets from that main sample and 31,429 from the control sample.
    
\end{enumerate}

\begin{table}[t]

\centering
\begin{tabular}{cccc}
\hline
Parameter & Range & Median & Units\\
\hline
\hline
Mass & $0.71-1.24$ & 1.05 &$M_\odot$\\
Radius & $0.71-1.65$ & 1.17 &$R_\odot$\\
log(g) & $4.04-4.63$ & 4.31& cgs\\
TESS Magnitude & $10.58-13.91$ & 13.09 & mag\\
Temperature & $4646-6376$ & 5819 & K\\
\hline
\end{tabular}
\caption{Stellar Sample Parameters: We show the 5$\%$ to 95$\%$ quantiles for each parameter range of our sample from the TIC \citep{Stassun2018}\label{tab:stellar}}
\end{table}

From these criteria, our main sample consisted
of 75,527 metal-poor stars that are below [Fe/H] $\leq$ -0.5 directly from the TIC. To increase the sample below [Fe/H] $\leq -0.5$ and better constrain the occurrence rate, we cross-matched the TIC with LAMOST DR8 \citep{Luo2015} (the TIC is only complete to LAMOST DR3) to identify stars with spectroscopic parameters derived after the creation of the TIC. From LAMOST DR8, we gained an additional 9,282 stars that pass our stellar sample cuts, resulting in a total of 84,809 metal-poor stars with metallicities below [Fe/H] $\leq-0.5$. 

 \begin{figure}
\begin{center}
\includegraphics[width=\linewidth]{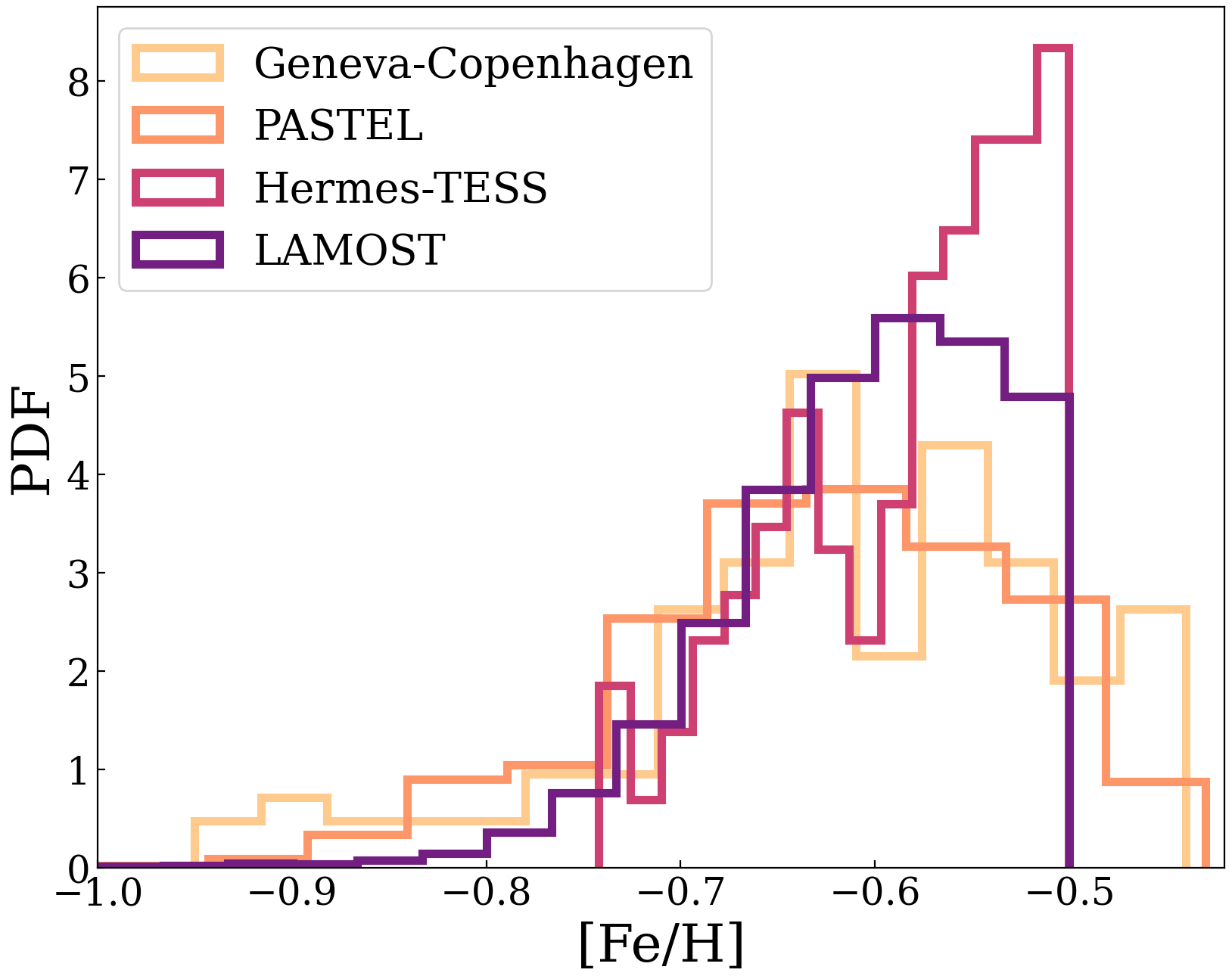}
\caption{The probability density function (PDF, \%) of the metallicities within our main sample ($-1\leq$[Fe/H]$ \leq -0.5$) by survey. 90\% of the stars within our sample have metallicities from LAMOST.\label{fig:Feh}}
\end{center}
\end{figure}

To enable a comparison with previous studies \citep[e.g.][]{Petigura2018,Zink2023}, we also included a control-sample of stars within the metallicity range of $-0.5 \leq$ [Fe/H] $\leq -0.25$. In doing so, we account for any systematic difference between the combined \emph{Kelper} and \emph{K2} occurrence rates and TESS. From the criteria listed above, our control sample consisted of 205,832 stars. Given the significant number of stars within this metallicity range, we included an additional cut to create our control sample, as the main objective of this study is to determine planet occurrence rates below [Fe/H] $ = -0.5$. We randomly selected 23,000 stars that produce a similar distribution of stellar properties as the metal-poor sample (see Figure \ref{fig:stellar}). With the addition of the moderately more metal-rich sub-sample, our total sample included 107,809 metal-poor stars. In Table \ref{tab:stellar}, we show the ranges and medians of the stellar properties for the total sample. 

From the TIC, we found that approximately 90\% of the stars within the final sample have metallicities from LAMOST, with the following surveys accounting for the remaining$\sim$10\%: PASTEL \citep{Soubiran2010}, Hermes-TESS \citep{Sharma2018}, and Geneva-Copenhagen \citep{Holmberg2009} (see Figure \ref{fig:Feh}). From these surveys, we determined the mean uncertainty for the metallicities to be $\sim0.04$dex. Given that the majority of the stars have metallicities derived from LAMOST spectroscopic data, the impact of systematic biases from other surveys is limited. However, for consistency, we transformed metallicities from PASTEL, Hermes-TESS, and Geneva-Copenhagen to LAMOST following the methodology in \citep{Soubiran2022}. In brief, we determined the offsets between LAMOST and each survey using targets that overlap with LAMOST. We found the overlaps between LAMOST to be 102, 61, and 24 stars for PASTEL, Hermes-TESS, and Geneva-Copenhagen, respectively. A linear fit was produced using the offsets for each survey and LAMOST, and the [Fe/H] measurements were then calibrated to the LAMOST system using the linear fit. We found offsets to be 0.12, 0.07, and 0.09 dex for PASTEL, Hermes-TESS, and Geneva-Copenhagen, respectively.

\begin{figure}[t!]
\begin{center}
\includegraphics[width=\linewidth]{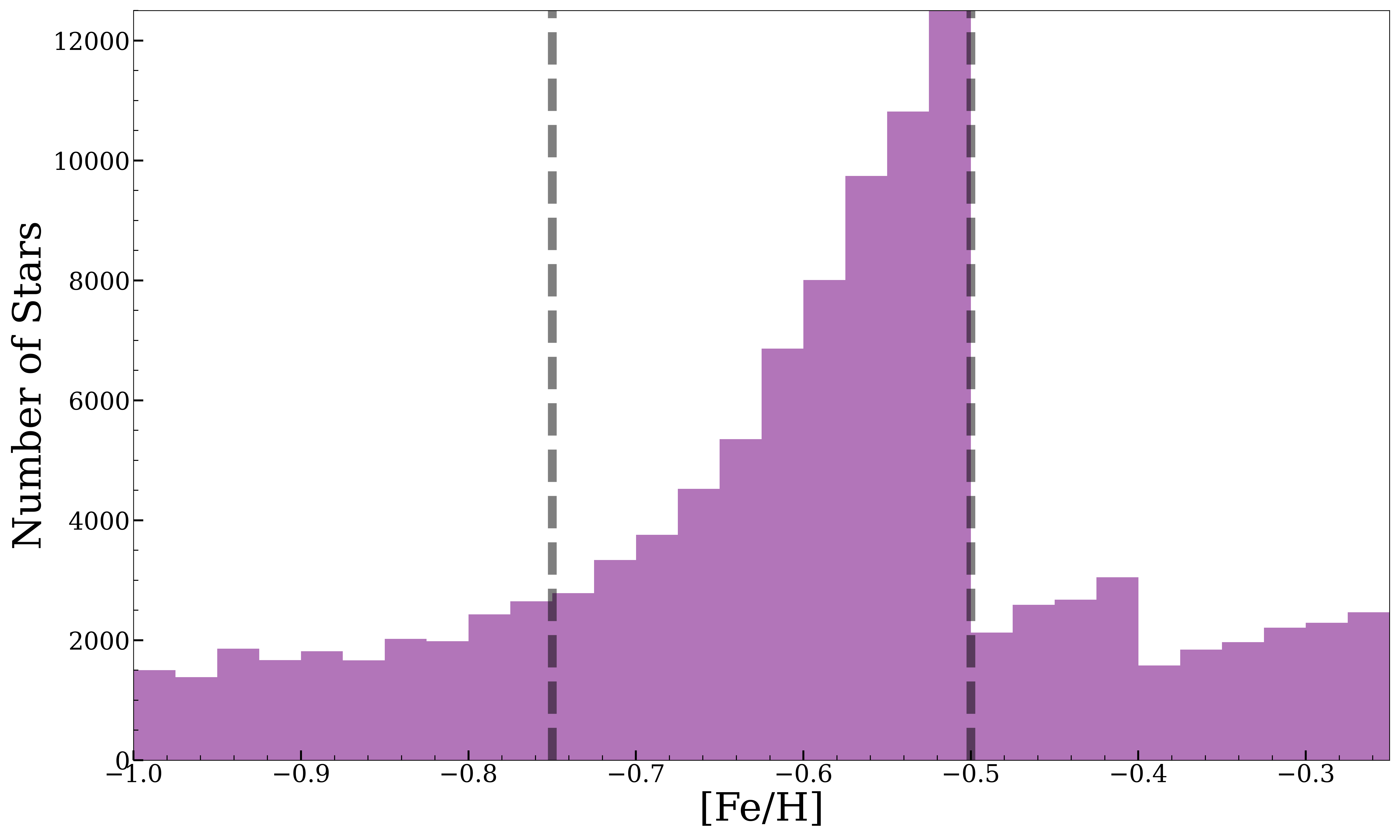}
\caption{We show a histogram of the stellar metallicities for our sample with bins indicated by black dashed lines. The sharp peak at [Fe/H]= -0.5 {is due to the reduction of the control sample as the main objective of this study is to determine super-Earth occurrence rates below [Fe/H]= -0.5} .\label{fig:FehHist}}
\end{center}
\end{figure}

In Figure \ref{fig:FehHist}, we show the stellar metallicities. The sharp peak at [Fe/H]= -0.5 is due to requiring spectroscopically derived metallicities [Fe/H]$\leq$ -0.5. Within the -0.25 to -0.5 bin, we note that the peak at approximately [Fe/H]= -0.4 results from the random cut performed when reducing the sample. 

\section{Planet Detection} \label{s:detection}
While \cite{Boley2021} describes our planet detection pipeline in detail, we briefly describe it here, noting additions and adjustments for this sample. {For each star in our sample,} we extract the light curves from TESS full-frame images (FFI). Before processing each light curve, we separated the data into the primary mission, 30-minute cadence observations (Sectors $<27$), and extended mission 1, 10-minute cadence (Sectors 27-55) observations. We do this to increase the detectability of planet candidates, as the primary and extended mission data do not have the same noise levels or systematics. Given that the primary and extended mission data are separated, we preferentially chose the 10-minute cadence data when they were available for a given target. The primary mission data were processed for targets without extended mission data.  

To optimize the light curves to be searched for planet candidates, we relied on the detrending software \texttt{wotan} \citep{wotan2019}. Within \texttt{wotan}, we employed Tukey's bi-weight method, which is indicated to be the most robust detrending method \citep{wotan2019}. Since the TESS data have sharp peaks at the beginning and ends of each observation sequence, we performed sigma-clipping to remove any systematics that \texttt{wotan} was unable to remove at the 5-$\sigma$ level. 

We used Transit Least Squares \citep[TLS, ][]{tls2009} as our transit search algorithm as it generally performs exceptionally well for smaller planets. Within TLS, we specified stellar mass, stellar radius, and limb-darkening coefficients, the latter of which were determined by interpolating the \citep{Claret2017} limb-darkening tables using the stellar effective temperature, surface gravity, and stellar metallicity from the TESS Input Catalog \citep{Stassun2018}. TLS produced a set of TCEs with a minimum of 3 transits that were then vetted and classified as planet candidates or false positives.

 \begin{figure}[t]
\begin{center}
\includegraphics[width=\linewidth]{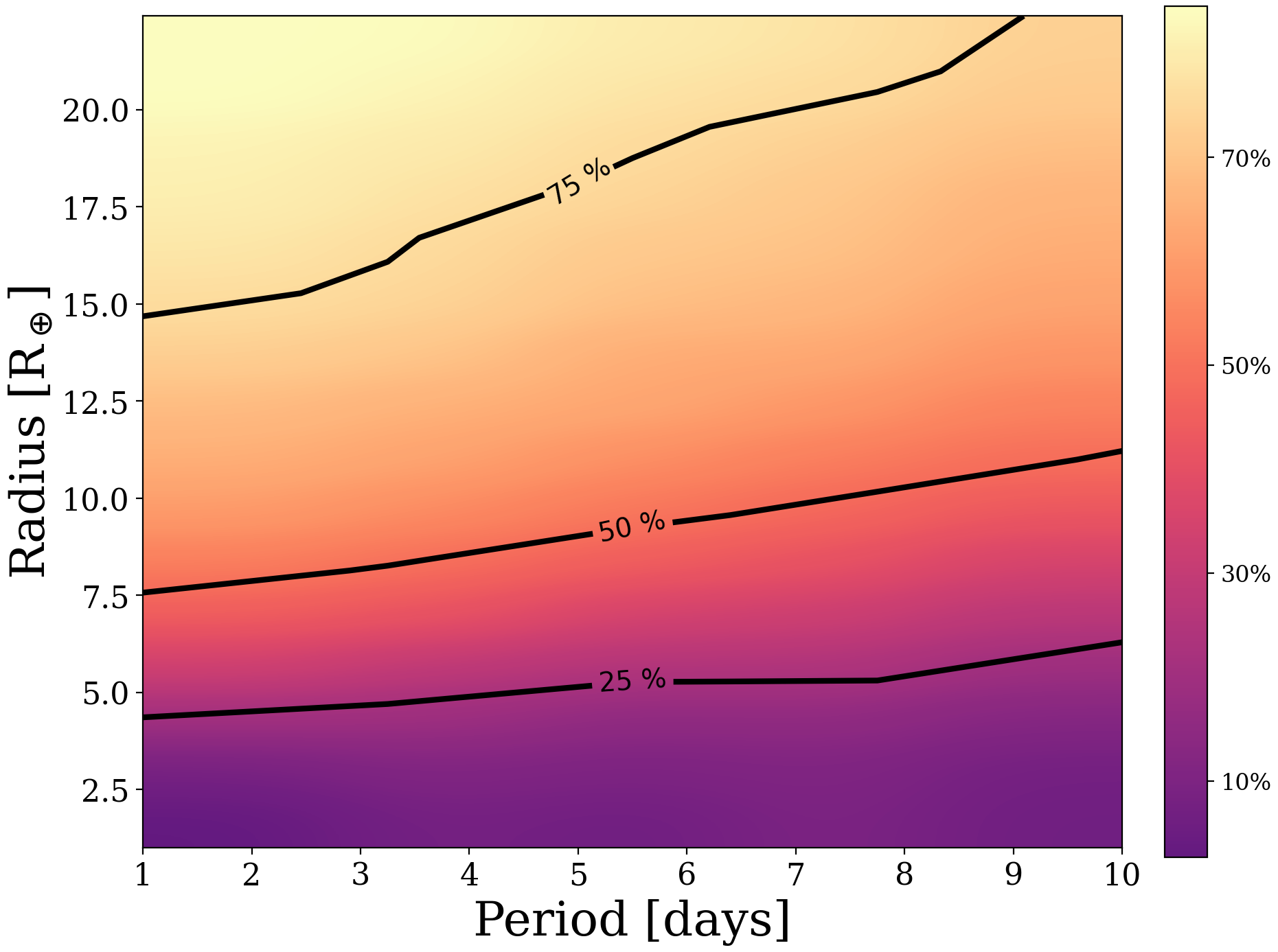}
\caption{We show the detection efficiency of our pipeline, which is calculated via injection-recovery test. Purple indicates a lower detection efficiency, whereas yellow indicates a higher detection efficiency. \label{fig:DE}}
\end{center}
\end{figure}

\subsection{Survey Completeness}
Using the framework in \cite{Boley2021}, we conducted injection and recovery tests to determine the detection efficiency of our pipeline and vetting process. { For each star in our sample,} we uniformly sampled for periods between 1--10 days and radii between 1--3~R$_{\oplus}$. We choose to focus on periods between 1--10 days to ensure a one-to-one comparison with previous studies \citep{Petigura2018,Zink2023}.
We assumed zero eccentricity, as planets on close-in orbits are subject to strong orbital circularization \citep{Montes2019}, and calculated limb-darkening coefficients as described above.

The simulated planets were injected into the light curves, which were then processed through our planet detection pipeline described above. We considered a simulated planet successfully recovered if the signal-to-noise ratio (S/N) was above 6$\sigma$, the false alarm probability was $\leq$ 0.0001, there were a minimum of 3 transits, and the period was within 1\% of the injected period. Using the injection and recovery tests, a detection efficiency grid was created in orbital period and planet radius space, shown in Figure \ref{fig:DE}. The survey completeness was then calculated by multiplying the detection efficiency and the geometric transit probability, which is the ratio of the combined host star and planet radius to the planet's orbital radius,

\begin{equation}\label{eq:transprob}
P_t(R_p,P)= \frac{R_p + R_\star}{a}
\end{equation}
where $R_\star$ is the stellar radius and \textit{a} the orbital semi-major
axis of the planet, determined from Kepler's third law.

\section{Forward Model} \label{s:forward}
Using the derived survey completeness, we can infer the intrinsic planet population by forward modeling. Given a population of planets, the forward model subjects the population to the selection effects of the detection pipeline, producing a simulated planet sample for comparison to the observed sample. Following previous studies \citep{Mulders2018,ExoMult,Zink2020A}, our forward modeling followed the procedure : 

\begin{enumerate}
    \item \textbf{Generate planet population}: A synthetic planet population was generated using a joint power-law in period-radius space and assigned a random orbital orientation;
    \item \textbf{Determine detected planets}: From the synthetic planet population, the number of transiting planets that would be observed by TESS was determined by simulating the instrumental and geometric selection effects; 
    \item \textbf{Calculate planet population likelihood}: Using a modified Poisson likelihood \citep{Zink2020A}, the detectable planet population was compared to the observed planet candidates to determine the goodness of fit; 
    \item \textbf{Repeat}: The likelihood continued to be drawn until the distribution of  parameters was well sampled, typically 500,000 iterations.
\end{enumerate}

To generate the planet population, we modeled the planet population distribution function as a joint power-law in planet radius (q(r)) and orbital period (g(p)) following the methodology outlined in \citep{ExoMult}. We assumed that the planet radius and orbital period distributions are independent, similar to previous studies, and modeled with broken power laws.

\begin{equation} 
\frac{d^2N}{drdp}=f q(r) g(p)
\end{equation}

\begin{equation} 
\begin{split}
q(r) &\propto  
\begin{cases}
r^{\alpha_1} & \text{if $r<R_{br}$} \\
r^{\alpha_2} & \text{if $r \geq R_{br}$} \\
\end{cases}\\
\end{split}
\end{equation}

\begin{equation} 
\begin{split}
g(p) &\propto
\begin{cases}
p^{\beta_1} & \text{if $p <P_{br}$} \\
p^{\beta_2} & \text{if $p \geq P_{br}$} \\
\end{cases} \\
\end{split}
\end{equation}
\\
where f is the number of planets per star within our occurrence model, $\alpha_1$, $\alpha_2$, $\beta_1$, and $\beta_2$ values are scaling model parameters. $P_{br}$ and $R_{br}$ represent the corresponding break in the period and radius power laws.


\begin{figure*}[t]
\begin{center}
\includegraphics[width=\linewidth]{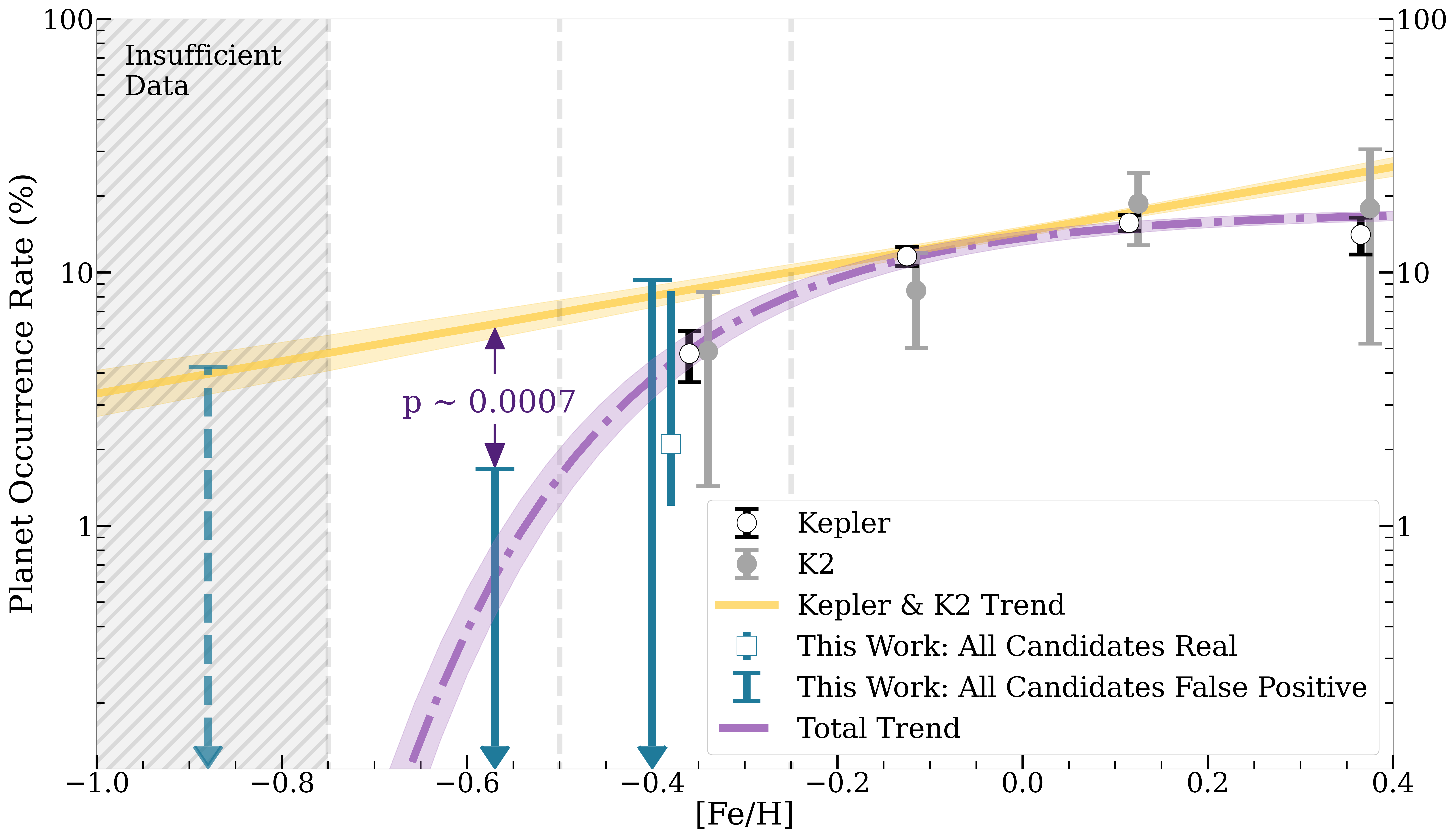}
\caption{Super-Earth occurrence for \emph{Kepler} (black circles) and \emph{K2} (gray circles), TESS data assuming all candidates are real (teal square), and 99.7\% confidence intervals to calculate the upper limits for TESS data assuming all candidates are false positives (teal triangles) as a function of metallicity. The [-0.25,-0.5] bin occurrence rates are offset horizontally for visual clarity. The best-fit power-law trend line for the \emph{Kepler} and \emph{K2} data (yellow) is displayed and extrapolated to [Fe/H]= -0.75 showing the 1-$\sigma$ uncertainties \citep{Zink2023}. We show the combined best-fit exponential trend line for \emph{Kepler}, \emph{K2}, and TESS (purple) (eq. \ref{eq:exp}) including the 1-$\sigma$ uncertainties. Each metallicity bin is indicated by gray dashed lines.  Within the [-0.75,-1] bin, there is insufficient data to further constrain the super-Earth occurrence rate as a function of metallicity (denoted by the gray hatched region).\label{fig:occur}}
\end{center}
\end{figure*}

\subsection{Model Optimization}
Using the simulated planet populations, we measured the Bayesian posterior for the seven model parameters by employing the emcee affine-invariant sampler \citep{Goodman2010,emcee2013} with 50 semi-independent walkers, 5000 burn-in steps, and 10,000 sample steps. We used uniform priors for all scaling parameters. Specifically, the priors for $\alpha_1$, $\alpha_2$, $\beta_1$, and $\beta_2$ range from -30 to +30. For $P_{br}$, we only considered the range from 1--10 days similar to the range of our sample. For $R_{br}$, the priors ranged from the minimum and maximum values for each planet type. Given the co-variance in radius between super-Earths and sub-Neptunes, we separated these planet populations along the population valley \citep{Fulton2017,VanEylen2018}. We used an empirically derived equation for the radius valley from \cite{Ho2023} (equation 4):

\begin{equation}
    log_{10}\left(\frac{R_p}{R_\oplus}\right)=m\hspace{1mm}log_{10}\left(\frac{P}{days}\right)+c
\end{equation}

where $m =-0.11$ and $c=0.37$.

\section{Results}\label{s:results}
Within our sample, we find a minimal increase in the detection efficiency as a function of metallicity. Stars at low metallicities are less active \cite{Amard2020}. They also have decreased opacities, resulting in smaller radii for a given temperature \citep{Xin2022}. Therefore, the difference in super-Earth detection efficiency between each metallicity bin ([$-0.25$, $-0.5$], [$-0.5$,$-0.75$], [$-0.75$,$-1.0$]) is $\sim$1\% on average, with the largest difference of 1.2\% between the [$-0.25$, $-0.5$] and [$-0.5$,$-0.75$] bins. Therefore, the average detection efficiency being $\sim$10\% in the [$-0.25$, $-0.5$] bin would be $\sim$11.2\% in the [$-0.5$,$-0.75$].

Figure \ref{fig:occur} displays the occurrence rates derived for our TESS sample, compared to previous analyses using \emph{Kepler} and \emph{K2} \citep{Zink2023}. We analyzed the  $-0.5 < [\mathrm{Fe/H}] \leq -0.25$ bin to overlap with those previous studies, to test for any systematic offsets from our TESS analysis. Within this bin, we found three planet candidates. We do not have any constraints on the false positive rate (i.e., reliability) of our TESS planet candidates, so we test the most optimistic (all planet candidates are real) and conservative (all planet candidates are false positives) scenarios. Assuming all candidates are real, we find $f_{real,[-0.25,-0.5]}=2.1^{+6.6}_{-0.99}\%$\footnote{The provided uncertainty is a direct measure of the 99.7\% confidence interval from the sampled occurrence rate ($f$) posterior distribution.}, consistent with the previous analyses. Assuming all the planet candidates are false positives, we place a 99.7\% confidence interval upper limit of $f_{fp,[-0.25,-0.5]}=9.32\%$. Our result is consistent with previous studies, which find the occurrence rates to be $4.78\pm 1.1\%$ $4.88\pm 3.45\%$ for \emph{Kepler} and \emph{K2}, respectively \citep{ Zink2023}. Therefore, we do not find any statistically significant systematic offsets between TESS and the \emph{Kepler} and \emph{K2} samples using our detection pipeline. Within the [$-0.5$,$-0.75$] bin, we detect no super-Earth candidates, and place a stringent 99.7\% confidence interval upper limit of $f_{[-0.5,-0.75]}=1.67\%$, shown in Figure \ref{fig:occur}. 

Similarly, no super-Earth candidates are detected in the [$-0.75$,$-1.0$] bin. We find  a 99.7\% confidence interval upper limit of $f_{[-0.75,-1]}=4.24\%$. However, given the limited sample size of 20,148 stars within this bin, we cannot further constrain super-Earth occurrence rate compared to \emph{Kepler} and \emph{K2} extrapolations (Figure \ref{fig:occur}). To determine whether in super-Earth occurrence rate would continue to decrease in the [$-0.75$,$-1.0$] bin would require a sample $\sim$ 4 times larger within that metallicity range based on our detection efficiency.

\subsection{Exponential fit} \label{s:exponential}
To determine an updated occurrence rate ($f_p$) trend as a function of metallicity, we combined the data from \emph{Kepler}, \emph{K2}, and TESS. We assume a simple exponential to model a cut-off in planet formation without introducing additional parameters above a power-law fit, of the form: 

\begin{equation}\label{eq:exp}
    f_p=f_0 \hspace{1mm} exp \left[ -10^{2\left([Fe/H]_{break}-[Fe/H]\right)} \right]
\end{equation}
where $f_0$ is the initial occurrence rate at $[Fe/H]=0.4$ and $[Fe/H]_{break}$ is the metallicity at which the slope of the exponential changes. To determine the best-fit parameters, we employed \texttt{scipy.optimize.curve$\_$fit}. This software relies on a nonlinear least squares method to fit the exponential \citep{Vugrin2007}.  We find the best-fit parameters to be $f_0 =17.33  \pm 0.7$ and $[Fe/H]_{break}= -0.31 \pm 0.02$ using 1$\sigma$ uncertainties.

\section{Discussion}\label{s:discuss}
A number of studies have considered planet occurrence rates as a function of metallicity \citep[see, e.g.][]{Petigura2018, Johnson2010,Udry2007, Fischer2005}. Recently, \cite{Zink2023} used \emph{Kepler} and \emph{K2} data to investigate the correlation between planet occurrence rates and metallicities for short-period (1--10 days) planets from 1--20~$R_{\oplus}$, directly comparable to our parameter space {that also considers short-period planet on orbits of 1--10 days} . Figure \ref{fig:occur} shows their results for \emph{Kepler} and \emph{K2}, and an extrapolation of their combined trend from both datasets to lower metallicities. Below [Fe/H] $\leq$ -0.5, the occurrence rate from the \emph{Kepler} and \emph{K2} extrapolation combined with our measured survey completeness should yield a detection of 68 super-Earth candidates in our sample of TESS stars with a total of $\sim$ 54 candidates in the [$-0.5$,$-0.75$] bin and $\sim$ 14 candidates in the [$-0.75$,$-1$]. {  To calculate the expected value of super-Earth candidates within our below [Fe/H] $\leq$ -0.5, we multiply the search completeness of our pipeline ($\sim$ 1.4\%) by the total sample ($\sim$ 85,000 stars). We then multiply that value by the average extrapolated \emph{Kepler} and \emph{K2} occurrence rate ($\sim$ 5.7\%) within that bin. Our new upper limit (1.67 \%) is well below the extrapolated trend ($\sim$ 6.06\%) within the [$-0.5$,$-0.75$] bin, and is statistically discrepant with a p-value = 0.000685. We calculate the p-value by calculating the overlap of the TESS and \emph{Kepler} and \emph{K2} distributions.} Our pipeline has a detection threshold of 6-$\sigma$, which is similar to detection thresholds used in previous TESS analyses \citep[e.g.,][]{Thuillier2022, Ment2023}. Adopting a considerably stricter requirement of 10-$\sigma$ for a TCE, reducing the parameter space in which we would detect signals but increasing the robustness of those signals, we find a lower but still strong discrepancy of with a p-value = 0.0293. This is strong evidence that the trend in \cite{Zink2023} \emph{cannot} be extended to lower metallicities, and that instead we are potentially seeing the onset of the expected critical metallicity cut-off for the formation of small planets. 

This discovery has implications for planet formation. First, it represents an incredibly useful, observational constraint on models of planet formation. The critical metallicity threshold implied here may already be in tension with, or rule out, a number of proposed models. By comparing the dust settling timescale to the metallicity-dependent disk lifetime \citep[e.g.,][]{Ercolano2010}, \citet{Johnson2012} derived the critical metallicity required to have enough solid material in the disk midplane before the dispersal of the disk (see their equation 10). At the orbital distances relevant for our sample, their critical metallicity corresponds to [Fe/H] = -2.5 which is two orders of magnitude lower than our [Fe/H]$_{break}$. Population synthesis models that require initial core formation near the ice line and large-scale migration predict a range of critical metallicity for small planet formation from [Fe/H] $>-1.8$ \citep{Hasegawa2014} to [Fe/H] $>-0.6$ \citep{Andama2024}, subject to disk parameters such as the level of turbulence. Similarly, using the pebble flux model of \citet{Lambrechts2014}, \citet{Lin2018} computed the core mass growth as a function of disk turbulence and metallicity (see their Figures 15 an 16). We have repeated their calculation to find that the formation of an Earth-mass core becomes difficult subject to disk turbulence when the metallicity falls below [Fe/H] $\sim$-0.7.
From direct numerical simulations, \citet{Li2021} find the critical metallicity to trigger the initial clumping of solids via streaming instability to be sensitively determined by the particle Stokes number, the disk radial pressure gradient, and turbulence. A solar or subsolar critical metallicity would generally require a large Stokes number reaching $\sim$0.1, which is near or above the maximum value expected in the analysis of nearby protoplanetary disks with concentric rings \citep[e.g.,][]{Rosotti20}. 

Second, it could imply that short-period super-Earths do not form early in the history of the universe. The vast majority of stars older than $\sim$7 billion years, nearly half the lifetime of the Galaxy, have metallicities below $-0.5$ \citep{Feuillet2019}. If small planet formation must wait until the Galaxy has been enriched to a third of solar metallicity or more \citep{Soubiran2008}, after the death of the initial generations of stars enriching the interstellar medium \citep{Abel2002,Beers2005,Frebel2007,Clark2008}, this threshold could directly inform the galactic inventory of small planets. 

Extending our analysis to stars of even lower metallicities could improve our understanding of the nature of the metallicity cut-off. Even constructing a larger sample within the [-0.5, -0.75] bin within this study would be instrumental in providing a strong constraint on the critical metallicity for planet formation and perhaps constraining the properties of the protoplanetary disk. Therefore, subsequent studies at metallicities below [Fe/H]=-0.75 will reveal the complete picture of planet formation across Galactic space and time.

\section{Summary \& Conclusions}\label{s:conclusions}
  
In this paper, we analyzed $\sim$110,000 stars with spectroscopically-derived metallicities within the metal-poor regime (-1 $\leq$ [Fe/H] $\leq$ -0.25). The objective of this study was to detect super-Earths at periods of 1--10 days and determine their occurrence rates in order to test theoretical predictions of the critical metallicity threshold for small planet formation \citep{Johnson2012,Lin2018,Li2021}. From this study, we present our main conclusions:

\begin{itemize} 
    \item We find a distinct metallicity “cliff” in for super-Earths at low metallicities (-0.75 $\leq$ [Fe/H] $\leq$ -0.5). This result suggests that super-Earths may be difficult to form within this regime, which, under certain disk conditions, is in line with formation theories that involve pebble accretion \citep[e.g.,][]{Lin2018}, planetesimal formation \citep[e.g.,][]{Andama2024} and more fundamentally the initial solid clumping by streaming instability \citep[e.g.,][]{Li2021} (\S \ref{s:results}).
    
    \item We find that planet occurrence rate trends above [Fe/H] $\gtrsim$ -0.5 likely cannot be extended to more metal-poor environments(\S \ref{s:results}). 

    \item We provide a functional form for super-Earth occurrence rates as a function of metallicity  (eq. \ref{eq:exp}, \S \ref{s:exponential}), and determine the metallicity cut-off to begin at $[Fe/H]_{break}= -0.31 \pm 0.02$ using 1$\sigma$ uncertainties.
\end{itemize}

Theory predicts that planet formation for all planet populations should become suppressed with decreasing metallicity \citep[e.g.,][]{Johnson2012, Lee2014,Lee2015}. However, this study is the first to discover empirical evidence suggesting that super-Earth formation may become significantly more difficult. Our study acts as an initial investigation into super-Earth formation in the metal-poor regime, but more studies are necessary to determine whether this metallicity cut-off may be as steep for longer period planets. Given that metal-poor stars have shorter disk lifetimes and smaller disk masses \citep[e.g.,][]{Yasui2010}, long-period planet formation may be suppressed as well. However, transit studies with longer baselines would be required to probe the metallicity-correlation for longer period super-Earths. These observations will likely be feasible with Roman and PLATO launch in the coming decade.

\newpage

\acknowledgments
We would like to thank Scott Gaudi for his insightful conversations and suggestions that have improved this work. K.M.B. acknowledges support from the NSF Graduate Research Fellowship Program under Grant No. (DGE1343012). This research has made use of the NASA Exoplanet Archive, which is operated by the California Institute of Technology, under contract with the National Aeronautics and Space Administration under the Exoplanet Exploration Program. This paper includes data collected by the TESS mission. Funding 
for the TESS mission is provided by the NASA's Science Mission Directorate. The results reported herein benefited from collaborations and/or information exchange within NASA’s Nexus for Exoplanet System Science (NExSS) research coordination network sponsored by NASA’s Science Mission Directorate. This material is based upon work supported by the National Aeronautics and Space Administration under Agreement No. 80NSSC21K0593 for the program “Alien Earths”. K.M.B. thanks the LSSTC Data Science Fellowship Program, which is funded by LSSTC, NSF Cybertraining Grant No. 1829740, the Brinson Foundation, and the Moore Foundation; her participation in the program has benefited this work.

 \software{ 
 \texttt{emcee}~\citep{emcee2013},\texttt{eleanor}~\citep{2019PASP..131i4502F}, \texttt{pylightcurve}~\citep{pylight},  \texttt{transit least quares algorithm}~ \citep{tls2009}, and \texttt{astropy} ~ \citep{Astropy2013}}

\clearpage

\bibliography{bibliography}

\begin{thebibliography}{}
\expandafter\ifx\csname natexlab\endcsname\relax\def\natexlab#1{#1}\fi
\providecommand{\url}[1]{\href{#1}{#1}}
\providecommand{\dodoi}[1]{doi:~\href{http://doi.org/#1}{\nolinkurl{#1}}}
\providecommand{\doeprint}[1]{\href{http://ascl.net/#1}{\nolinkurl{http://ascl.net/#1}}}
\providecommand{\doarXiv}[1]{\href{https://arxiv.org/abs/#1}{\nolinkurl{https://arxiv.org/abs/#1}}}

\bibitem[{{Abel} {et~al.}(2002){Abel}, {Bryan}, \& {Norman}}]{Abel2002}
{Abel}, T., {Bryan}, G.~L., \& {Norman}, M.~L. 2002, Science, 295, 93, \dodoi{10.1126/science.295.5552.93}

\bibitem[{{Adibekyan} {et~al.}(2012{\natexlab{a}}){Adibekyan}, {Delgado Mena}, {Sousa}, {Santos}, {Israelian}, {Gonz{\'a}lez Hern{\'a}ndez}, {Mayor}, \& {Hakobyan}}]{Adibekyan2012}
{Adibekyan}, V.~Z., {Delgado Mena}, E., {Sousa}, S.~G., {et~al.} 2012{\natexlab{a}}, A\&A, 547, A36, \dodoi{10.1051/0004-6361/201220167}

\bibitem[{{Adibekyan} {et~al.}(2012{\natexlab{b}}){Adibekyan}, {Santos}, {Sousa}, {Israelian}, {Delgado Mena}, {Gonz{\'a}lez Hern{\'a}ndez}, {Mayor}, {Lovis}, \& {Udry}}]{Adibekyan2012a}
{Adibekyan}, V.~Z., {Santos}, N.~C., {Sousa}, S.~G., {et~al.} 2012{\natexlab{b}}, A\&A, 543, A89, \dodoi{10.1051/0004-6361/201219564}

\bibitem[{{Akeson} {et~al.}(2013){Akeson}, {Chen}, {Ciardi}, {Crane}, {Good}, {Harbut}, {Jackson}, {Kane}, {Laity}, {Leifer}, {Lynn}, {McElroy}, {Papin}, {Plavchan}, {Ram{\'\i}rez}, {Rey}, {von Braun}, {Wittman}, {Abajian}, {Ali}, {Beichman}, {Beekley}, {Berriman}, {Berukoff}, {Bryden}, {Chan}, {Groom}, {Lau}, {Payne}, {Regelson}, {Saucedo}, {Schmitz}, {Stauffer}, {Wyatt}, \& {Zhang}}]{Exoarchive2013}
{Akeson}, R.~L., {Chen}, X., {Ciardi}, D., {et~al.} 2013, PASP, 125, 989, \dodoi{10.1086/672273}

\bibitem[{{Alvarado-Montes} \& {Garc{\'\i}a-Carmona}(2019)}]{Montes2019}
{Alvarado-Montes}, J.~A., \& {Garc{\'\i}a-Carmona}, C. 2019, MNRAS, 486, 3963, \dodoi{10.1093/mnras/stz1081}

\bibitem[{{Amard} \& {Matt}(2020)}]{Amard2020}
{Amard}, L., \& {Matt}, S.~P. 2020, \apj, 889, 108, \dodoi{10.3847/1538-4357/ab6173}

\bibitem[{{Andama} {et~al.}(2024){Andama}, {Mah}, \& {Bitsch}}]{Andama2024}
{Andama}, G., {Mah}, J., \& {Bitsch}, B. 2024, arXiv e-prints, arXiv:2401.16155, \dodoi{10.48550/arXiv.2401.16155}

\bibitem[{Anderson \& Darling(1952)}]{Anderson1952}
Anderson, T.~W., \& Darling, D.~A. 1952, The Annals of Mathematical Statistics, 23, 193.
\newblock \url{http://www.jstor.org/stable/2236446}

\bibitem[{{Angelov}(1996)}]{Angelov1996}
{Angelov}, T. 1996, Bulletin Astronomique de Belgrade, 154, 13

\bibitem[{{Arriagada}(2011)}]{Arriagada2011}
{Arriagada}, P. 2011, ApJ, 734, 70, \dodoi{10.1088/0004-637X/734/1/70}

\bibitem[{{Astropy Collaboration} {et~al.}(2013){Astropy Collaboration}, {Robitaille}, {Tollerud}, {Greenfield}, {Droettboom}, {Bray}, {Aldcroft}, {Davis}, {Ginsburg}, {Price-Whelan}, {Kerzendorf}, {Conley}, {Crighton}, {Barbary}, {Muna}, {Ferguson}, {Grollier}, {Parikh}, {Nair}, {Unther}, {Deil}, {Woillez}, {Conseil}, {Kramer}, {Turner}, {Singer}, {Fox}, {Weaver}, {Zabalza}, {Edwards}, {Azalee Bostroem}, {Burke}, {Casey}, {Crawford}, {Dencheva}, {Ely}, {Jenness}, {Labrie}, {Lim}, {Pierfederici}, {Pontzen}, {Ptak}, {Refsdal}, {Servillat}, \& {Streicher}}]{Astropy2013}
{Astropy Collaboration}, {Robitaille}, T.~P., {Tollerud}, E.~J., {et~al.} 2013, A\&A, 558, A33, \dodoi{10.1051/0004-6361/201322068}

\bibitem[{{Bai} \& {Stone}(2010)}]{Bai2010}
{Bai}, X.-N., \& {Stone}, J.~M. 2010, ApJ, 722, 1437, \dodoi{10.1088/0004-637X/722/2/1437}

\bibitem[{{Bailey} \& {Batygin}(2018)}]{Bailey2018}
{Bailey}, E., \& {Batygin}, K. 2018, ApJL, 866, L2, \dodoi{10.3847/2041-8213/aade90}

\bibitem[{{Batygin} \& {Brown}(2016)}]{Batygin2016}
{Batygin}, K., \& {Brown}, M.~E. 2016, AJ, 151, 22, \dodoi{10.3847/0004-6256/151/2/22}

\bibitem[{{Batygin} \& {Morbidelli}(2023)}]{Batygin2023}
{Batygin}, K., \& {Morbidelli}, A. 2023, Nature Astronomy, 7, 330, \dodoi{10.1038/s41550-022-01850-5}

\bibitem[{{Beers} \& {Christlieb}(2005)}]{Beers2005}
{Beers}, T.~C., \& {Christlieb}, N. 2005, AAR\&A, 43, 531, \dodoi{10.1146/annurev.astro.42.053102.134057}

\bibitem[{{Bernstein} {et~al.}(2003){Bernstein}, {Shectman}, {Gunnels}, {Mochnacki}, \& {Athey}}]{Bernstein03}
{Bernstein}, R., {Shectman}, S.~A., {Gunnels}, S.~M., {Mochnacki}, S., \& {Athey}, A.~E. 2003, Proc. SPIE, 4841, 1694, \dodoi{10.1117/12.461502}

\bibitem[{{Boley} {et~al.}(2021){Boley}, {Wang}, {Zinn}, {Collins}, {Collins}, {Gan}, \& {Li}}]{Boley2021}
{Boley}, K.~M., {Wang}, J., {Zinn}, J.~C., {et~al.} 2021, AJ, 162, 85, \dodoi{10.3847/1538-3881/ac0e2d}

\bibitem[{{Bromm} {et~al.}(2001){Bromm}, {Ferrara}, {Coppi}, \& {Larson}}]{Bromm2001}
{Bromm}, V., {Ferrara}, A., {Coppi}, P.~S., \& {Larson}, R.~B. 2001, MNRAS, 328, 969, \dodoi{10.1046/j.1365-8711.2001.04915.x}

\bibitem[{{Burgasser} {et~al.}(2003){Burgasser}, {Kirkpatrick}, {Reid}, {Brown}, {Miskey}, \& {Gizis}}]{Burgasser2003}
{Burgasser}, A.~J., {Kirkpatrick}, J.~D., {Reid}, I.~N., {et~al.} 2003, ApJ, 586, 512, \dodoi{10.1086/346263}

\bibitem[{{Burrows} {et~al.}(1993){Burrows}, {Hubbard}, {Saumon}, \& {Lunine}}]{Burrows1993}
{Burrows}, A., {Hubbard}, W.~B., {Saumon}, D., \& {Lunine}, J.~I. 1993, \apj, 406, 158, \dodoi{10.1086/172427}

\bibitem[{{Chen} \& {Kipping}(2017)}]{Chen2017}
{Chen}, J., \& {Kipping}, D. 2017, ApJ, 834, 17, \dodoi{10.3847/1538-4357/834/1/17}

\bibitem[{{Choi} \& {Nagamine}(2009)}]{Choi2009}
{Choi}, J.-H., \& {Nagamine}, K. 2009, MNRAS, 393, 1595, \dodoi{10.1111/j.1365-2966.2008.14297.x}

\bibitem[{{Claret}(2017)}]{Claret2017}
{Claret}, A. 2017, A\&A, 600, A30, \dodoi{10.1051/0004-6361/201629705}

\bibitem[{{Claret} \& {Bloemen}(2011)}]{claret}
{Claret}, A., \& {Bloemen}, S. 2011, A\&A, 529, A75, \dodoi{10.1051/0004-6361/201116451}

\bibitem[{{Clark} {et~al.}(2008){Clark}, {Glover}, \& {Klessen}}]{Clark2008}
{Clark}, P.~C., {Glover}, S. C.~O., \& {Klessen}, R.~S. 2008, ApJ, 672, 757, \dodoi{10.1086/524187}

\bibitem[{{Creevey} {et~al.}(2023){Creevey}, {Sordo}, {Pailler}, {Fr{\'e}mat}, {Heiter}, {Th{\'e}venin}, {Andrae}, {Fouesneau}, {Lobel}, {Bailer-Jones}, {Garabato}, {Bellas-Velidis}, {Brugaletta}, {Lorca}, {Ordenovic}, {Palicio}, {Sarro}, {Delchambre}, {Drimmel}, {Rybizki}, {Torralba Elipe}, {Korn}, {Recio-Blanco}, {Schultheis}, {De Angeli}, {Montegriffo}, {Abreu Aramburu}, {Accart}, {{\'A}lvarez}, {Bakker}, {Brouillet}, {Burlacu}, {Carballo}, {Casamiquela}, {Chiavassa}, {Contursi}, {Cooper}, {Dafonte}, {Dapergolas}, {de Laverny}, {Dharmawardena}, {Edvardsson}, {Le Fustec}, {Garc{\'\i}a-Lario}, {Garc{\'\i}a-Torres}, {Gomez}, {Gonz{\'a}lez-Santamar{\'\i}a}, {Hatzidimitriou}, {Jean-Antoine Piccolo}, {Kontiza}, {Kordopatis}, {Lanzafame}, {Lebreton}, {Licata}, {Lindstr{\o}m}, {Livanou}, {Magdaleno Romeo}, {Manteiga}, {Marocco}, {Marshall}, {Mary}, {Nicolas}, {Pallas-Quintela}, {Panem}, {Pichon}, {Poggio}, {Riclet}, {Robin}, {Santove{\~n}a}, {Silvelo}, {Slezak}, {Smart}, {Soubiran}, {S{\"u}veges}, {Ulla},
  {Utrilla}, {Vallenari}, {Zhao}, {Zorec}, {Barrado}, {Bijaoui}, {Bouret}, {Blomme}, {Brott}, {Cassisi}, {Kochukhov}, {Martayan}, {Shulyak}, \& {Silvester}}]{GaiaDR3}
{Creevey}, O.~L., {Sordo}, R., {Pailler}, F., {et~al.} 2023, A\&A, 674, A26, \dodoi{10.1051/0004-6361/202243688}

\bibitem[{{Cumming} {et~al.}(1999){Cumming}, {Marcy}, \& {Butler}}]{Cumming1999}
{Cumming}, A., {Marcy}, G.~W., \& {Butler}, R.~P. 1999, ApJ, 526, 890, \dodoi{10.1086/308020}

\bibitem[{Dressing \& Charbonneau(2015)}]{Dressing2015}
Dressing, C.~D., \& Charbonneau, D. 2015, The Astrophysical Journal, 807, 45, \dodoi{10.1088/0004-637x/807/1/45}

\bibitem[{{El-Badry} \& {Rix}(2019)}]{El-Badry2019}
{El-Badry}, K., \& {Rix}, H.-W. 2019, \mnras, 482, L139, \dodoi{10.1093/mnrasl/sly206}

\bibitem[{{Endl} {et~al.}(2006{\natexlab{a}}){Endl}, {Cochran}, {K{\"u}rster}, {Paulson}, {Wittenmyer}, {MacQueen}, \& {Tull}}]{2006ApJ...649..436E}
{Endl}, M., {Cochran}, W.~D., {K{\"u}rster}, M., {et~al.} 2006{\natexlab{a}}, ApJ, 649, 436, \dodoi{10.1086/506465}

\bibitem[{{Endl} {et~al.}(2006{\natexlab{b}}){Endl}, {Cochran}, {K{\"u}rster}, {Paulson}, {Wittenmyer}, {MacQueen}, \& {Tull}}]{Endl2006}
---. 2006{\natexlab{b}}, ApJ, 649, 436, \dodoi{10.1086/506465}

\bibitem[{{Ercolano} \& {Clarke}(2010)}]{Ercolano2010}
{Ercolano}, B., \& {Clarke}, C.~J. 2010, MNRAS, 402, 2735, \dodoi{10.1111/j.1365-2966.2009.16094.x}

\bibitem[{{Feinstein} {et~al.}(2019){Feinstein}, {Montet}, {Foreman-Mackey}, {Bedell}, {Saunders}, {Bean}, {Christiansen}, {Hedges}, {Luger}, {Scolnic}, \& {Cardoso}}]{2019PASP..131i4502F}
{Feinstein}, A.~D., {Montet}, B.~T., {Foreman-Mackey}, D., {et~al.} 2019, PASP, 131, 094502, \dodoi{10.1088/1538-3873/ab291c}

\bibitem[{{Feuillet} {et~al.}(2019){Feuillet}, {Frankel}, {Lind}, {Frinchaboy}, {Garc{\'\i}a-Hern{\'a}ndez}, {Lane}, {Nitschelm}, \& {Roman-Lopes}}]{Feuillet2019}
{Feuillet}, D.~K., {Frankel}, N., {Lind}, K., {et~al.} 2019, \mnras, 489, 1742, \dodoi{10.1093/mnras/stz2221}

\bibitem[{{Fischer} \& {Valenti}(2005)}]{Fischer2005}
{Fischer}, D.~A., \& {Valenti}, J. 2005, ApJ, 622, 1102, \dodoi{10.1086/428383}

\bibitem[{{Foreman-Mackey} {et~al.}(2013){Foreman-Mackey}, {Hogg}, {Lang}, \& {Goodman}}]{emcee2013}
{Foreman-Mackey}, D., {Hogg}, D.~W., {Lang}, D., \& {Goodman}, J. 2013, PASP, 125, 306, \dodoi{10.1086/670067}

\bibitem[{{Forgan} {et~al.}(2017){Forgan}, {Rowlands}, {Gomez}, {Gomez}, {Schofield}, {Dunne}, \& {Maddox}}]{Forgan2017}
{Forgan}, D.~H., {Rowlands}, K., {Gomez}, H.~L., {et~al.} 2017, MNRAS, 472, 2289, \dodoi{10.1093/mnras/stx2162}

\bibitem[{{Frebel} {et~al.}(2007){Frebel}, {Johnson}, \& {Bromm}}]{Frebel2007}
{Frebel}, A., {Johnson}, J.~L., \& {Bromm}, V. 2007, MNRAS, 380, L40, \dodoi{10.1111/j.1745-3933.2007.00344.x}

\bibitem[{{Fressin} {et~al.}(2013){Fressin}, {Torres}, {Charbonneau}, {Bryson}, {Christiansen}, {Dressing}, {Jenkins}, {Walkowicz}, \& {Batalha}}]{Fressin2013}
{Fressin}, F., {Torres}, G., {Charbonneau}, D., {et~al.} 2013, ApJ, 766, 81, \dodoi{10.1088/0004-637X/766/2/81}

\bibitem[{{Fulton} \& {Petigura}(2018)}]{Fulton2018}
{Fulton}, B.~J., \& {Petigura}, E.~A. 2018, AJ, 156, 264, \dodoi{10.3847/1538-3881/aae828}

\bibitem[{{Fulton} {et~al.}(2017){Fulton}, {Petigura}, {Howard}, {Isaacson}, {Marcy}, {Cargile}, {Hebb}, {Weiss}, {Johnson}, {Morton}, {Sinukoff}, {Crossfield}, \& {Hirsch}}]{Fulton2017}
{Fulton}, B.~J., {Petigura}, E.~A., {Howard}, A.~W., {et~al.} 2017, AJ, 154, 109, \dodoi{10.3847/1538-3881/aa80eb}

\bibitem[{{Fung} \& {Lee}(2018)}]{Fung2018}
{Fung}, J., \& {Lee}, E.~J. 2018, \apj, 859, 126, \dodoi{10.3847/1538-4357/aabaf7}

\bibitem[{{Gaia Collaboration} {et~al.}(2018){Gaia Collaboration}, {Brown}, {Vallenari}, {Prusti}, {de Bruijne}, {Babusiaux}, {Bailer-Jones}, {Biermann}, {Evans}, {Eyer}, {Jansen}, {Jordi}, {Klioner}, {Lammers}, {Lindegren}, {Luri}, {Mignard}, {Panem}, {Pourbaix}, {Randich}, {Sartoretti}, {Siddiqui}, {Soubiran}, {van Leeuwen}, {Walton}, {Arenou}, {Bastian}, {Cropper}, {Drimmel}, {Katz}, {Lattanzi}, {Bakker}, {Cacciari}, {Casta{\~n}eda}, {Chaoul}, {Cheek}, {De Angeli}, {Fabricius}, {Guerra}, {Holl}, {Masana}, {Messineo}, {Mowlavi}, {Nienartowicz}, {Panuzzo}, {Portell}, {Riello}, {Seabroke}, {Tanga}, {Th{\'e}venin}, {Gracia-Abril}, {Comoretto}, {Garcia-Reinaldos}, {Teyssier}, {Altmann}, {Andrae}, {Audard}, {Bellas-Velidis}, {Benson}, {Berthier}, {Blomme}, {Burgess}, {Busso}, {Carry}, {Cellino}, {Clementini}, {Clotet}, {Creevey}, {Davidson}, {De Ridder}, {Delchambre}, {Dell'Oro}, {Ducourant}, {Fern{\'a}ndez-Hern{\'a}ndez}, {Fouesneau}, {Fr{\'e}mat}, {Galluccio}, {Garc{\'\i}a-Torres},
  {Gonz{\'a}lez-N{\'u}{\~n}ez}, {Gonz{\'a}lez-Vidal}, {Gosset}, {Guy}, {Halbwachs}, {Hambly}, {Harrison}, {Hern{\'a}ndez}, {Hestroffer}, {Hodgkin}, {Hutton}, {Jasniewicz}, {Jean-Antoine-Piccolo}, {Jordan}, {Korn}, {Krone-Martins}, {Lanzafame}, {Lebzelter}, {L{\"o}ffler}, {Manteiga}, {Marrese}, {Mart{\'\i}n-Fleitas}, {Moitinho}, {Mora}, {Muinonen}, {Osinde}, {Pancino}, {Pauwels}, {Petit}, {Recio-Blanco}, {Richards}, {Rimoldini}, {Robin}, {Sarro}, {Siopis}, {Smith}, {Sozzetti}, {S{\"u}veges}, {Torra}, {van Reeven}, {Abbas}, {Abreu Aramburu}, {Accart}, {Aerts}, {Altavilla}, {{\'A}lvarez}, {Alvarez}, {Alves}, {Anderson}, {Andrei}, {Anglada Varela}, {Antiche}, {Antoja}, {Arcay}, {Astraatmadja}, {Bach}, {Baker}, {Balaguer-N{\'u}{\~n}ez}, {Balm}, {Barache}, {Barata}, {Barbato}, {Barblan}, {Barklem}, {Barrado}, {Barros}, {Barstow}, {Bartholom{\'e} Mu{\~n}oz}, {Bassilana}, {Becciani}, {Bellazzini}, {Berihuete}, {Bertone}, {Bianchi}, {Bienaym{\'e}}, {Blanco-Cuaresma}, {Boch}, {Boeche}, {Bombrun}, {Borrachero},
  {Bossini}, {Bouquillon}, {Bourda}, {Bragaglia}, {Bramante}, {Breddels}, {Bressan}, {Brouillet}, {Br{\"u}semeister}, {Brugaletta}, {Bucciarelli}, {Burlacu}, {Busonero}, {Butkevich}, {Buzzi}, {Caffau}, {Cancelliere}, {Cannizzaro}, {Cantat-Gaudin}, {Carballo}, {Carlucci}, {Carrasco}, {Casamiquela}, {Castellani}, {Castro-Ginard}, {Charlot}, {Chemin}, {Chiavassa}, {Cocozza}, {Costigan}, {Cowell}, {Crifo}, {Crosta}, {Crowley}, {Cuypers}, {Dafonte}, {Damerdji}, {Dapergolas}, {David}, {David}, {de Laverny}, {De Luise}, {De March}, {de Martino}, {de Souza}, {de Torres}, {Debosscher}, {del Pozo}, {Delbo}, {Delgado}, {Delgado}, {Di Matteo}, {Diakite}, {Diener}, {Distefano}, {Dolding}, {Drazinos}, {Dur{\'a}n}, {Edvardsson}, {Enke}, {Eriksson}, {Esquej}, {Eynard Bontemps}, {Fabre}, {Fabrizio}, {Faigler}, {Falc{\~a}o}, {Farr{\`a}s Casas}, {Federici}, {Fedorets}, {Fernique}, {Figueras}, {Filippi}, {Findeisen}, {Fonti}, {Fraile}, {Fraser}, {Fr{\'e}zouls}, {Gai}, {Galleti}, {Garabato}, {Garc{\'\i}a-Sedano}, {Garofalo},
  {Garralda}, {Gavel}, {Gavras}, {Gerssen}, {Geyer}, {Giacobbe}, {Gilmore}, {Girona}, {Giuffrida}, {Glass}, {Gomes}, {Granvik}, {Gueguen}, {Guerrier}, {Guiraud}, {Guti{\'e}rrez-S{\'a}nchez}, {Haigron}, {Hatzidimitriou}, {Hauser}, {Haywood}, {Heiter}, {Helmi}, {Heu}, {Hilger}, {Hobbs}, {Hofmann}, {Holland}, {Huckle}, {Hypki}, {Icardi}, {Jan{\ss}en}, {Jevardat de Fombelle}, {Jonker}, {Juh{\'a}sz}, {Julbe}, {Karampelas}, {Kewley}, {Klar}, {Kochoska}, {Kohley}, {Kolenberg}, {Kontizas}, {Kontizas}, {Koposov}, {Kordopatis}, {Kostrzewa-Rutkowska}, {Koubsky}, {Lambert}, {Lanza}, {Lasne}, {Lavigne}, {Le Fustec}, {Le Poncin-Lafitte}, {Lebreton}, {Leccia}, {Leclerc}, {Lecoeur-Taibi}, {Lenhardt}, {Leroux}, {Liao}, {Licata}, {Lindstr{\o}m}, {Lister}, {Livanou}, {Lobel}, {L{\'o}pez}, {Managau}, {Mann}, {Mantelet}, {Marchal}, {Marchant}, {Marconi}, {Marinoni}, {Marschalk{\'o}}, {Marshall}, {Martino}, {Marton}, {Mary}, {Massari}, {Matijevi{\v{c}}}, {Mazeh}, {McMillan}, {Messina}, {Michalik}, {Millar}, {Molina}, {Molinaro},
  {Moln{\'a}r}, {Montegriffo}, {Mor}, {Morbidelli}, {Morel}, {Morris}, {Mulone}, {Muraveva}, {Musella}, {Nelemans}, {Nicastro}, {Noval}, {O'Mullane}, {Ord{\'e}novic}, {Ord{\'o}{\~n}ez-Blanco}, {Osborne}, {Pagani}, {Pagano}, {Pailler}, {Palacin}, {Palaversa}, {Panahi}, {Pawlak}, {Piersimoni}, {Pineau}, {Plachy}, {Plum}, {Poggio}, {Poujoulet}, {Pr{\v{s}}a}, {Pulone}, {Racero}, {Ragaini}, {Rambaux}, {Ramos-Lerate}, {Regibo}, {Reyl{\'e}}, {Riclet}, {Ripepi}, {Riva}, {Rivard}, {Rixon}, {Roegiers}, {Roelens}, {Romero-G{\'o}mez}, {Rowell}, {Royer}, {Ruiz-Dern}, {Sadowski}, {Sagrist{\`a} Sell{\'e}s}, {Sahlmann}, {Salgado}, {Salguero}, {Sanna}, {Santana-Ros}, {Sarasso}, {Savietto}, {Schultheis}, {Sciacca}, {Segol}, {Segovia}, {S{\'e}gransan}, {Shih}, {Siltala}, {Silva}, {Smart}, {Smith}, {Solano}, {Solitro}, {Sordo}, {Soria Nieto}, {Souchay}, {Spagna}, {Spoto}, {Stampa}, {Steele}, {Steidelm{\"u}ller}, {Stephenson}, {Stoev}, {Suess}, {Surdej}, {Szabados}, {Szegedi-Elek}, {Tapiador}, {Taris}, {Tauran}, {Taylor},
  {Teixeira}, {Terrett}, {Teyssand ier}, {Thuillot}, {Titarenko}, {Torra Clotet}, {Turon}, {Ulla}, {Utrilla}, {Uzzi}, {Vaillant}, {Valentini}, {Valette}, {van Elteren}, {Van Hemelryck}, {van Leeuwen}, {Vaschetto}, {Vecchiato}, {Veljanoski}, {Viala}, {Vicente}, {Vogt}, {von Essen}, {Voss}, {Votruba}, {Voutsinas}, {Walmsley}, {Weiler}, {Wertz}, {Wevers}, {Wyrzykowski}, {Yoldas}, {{\v{Z}}erjal}, {Ziaeepour}, {Zorec}, {Zschocke}, {Zucker}, {Zurbach}, \& {Zwitter}}]{Gaia2018}
{Gaia Collaboration}, {Brown}, A.~G.~A., {Vallenari}, A., {et~al.} 2018, A\&A, 616, A1, \dodoi{10.1051/0004-6361/201833051}

\bibitem[{{Goldreich} {et~al.}(2004){Goldreich}, {Lithwick}, \& {Sari}}]{Goldreich2004}
{Goldreich}, P., {Lithwick}, Y., \& {Sari}, R. 2004, \araa, 42, 549, \dodoi{10.1146/annurev.astro.42.053102.134004}

\bibitem[{{Gonzalez} {et~al.}(2001){Gonzalez}, {Laws}, {Tyagi}, \& {Reddy}}]{gon1997}
{Gonzalez}, G., {Laws}, C., {Tyagi}, S., \& {Reddy}, B.~E. 2001, AJ, 121, 432, \dodoi{10.1086/318048}

\bibitem[{{Goodman} \& {Weare}(2010)}]{Goodman2010}
{Goodman}, J., \& {Weare}, J. 2010, Communications in Applied Mathematics and Computational Science, 5, 65, \dodoi{10.2140/camcos.2010.5.65}

\bibitem[{{Guo} {et~al.}(2017){Guo}, {Johnson}, {Mann}, {Kraus}, {Curtis}, \& {Latham}}]{Guo2017}
{Guo}, X., {Johnson}, J.~A., {Mann}, A.~W., {et~al.} 2017, ApJ, 838, 25, \dodoi{10.3847/1538-4357/aa6004}

\bibitem[{{Hartman} \& {Bakos}(2016)}]{Vartools2016}
{Hartman}, J.~D., \& {Bakos}, G.~{\'A}. 2016, Astronomy and Computing, 17, 1, \dodoi{10.1016/j.ascom.2016.05.006}

\bibitem[{{Hasegawa} \& {Hirashita}(2014)}]{Hasegawa2014}
{Hasegawa}, Y., \& {Hirashita}, H. 2014, ApJ, 788, 62, \dodoi{10.1088/0004-637X/788/1/62}

\bibitem[{{Haworth} {et~al.}(2016){Haworth}, {Ilee}, {Forgan}, {Facchini}, {Price}, {Boneberg}, {Booth}, {Clarke}, {Gonzalez}, {Hutchison}, {Kamp}, {Laibe}, {Lyra}, {Meru}, {Mohanty}, {Pani{\'c}}, {Rice}, {Suzuki}, {Teague}, {Walsh}, {Woitke}, \& {Community authors}}]{Haworth2016}
{Haworth}, T.~J., {Ilee}, J.~D., {Forgan}, D.~H., {et~al.} 2016, \pasa, 33, e053, \dodoi{10.1017/pasa.2016.45}

\bibitem[{{Haywood}(2008)}]{Haywood2008}
{Haywood}, M. 2008, A\&A, 482, 673, \dodoi{10.1051/0004-6361:20079141}

\bibitem[{{Hellier} {et~al.}(2014){Hellier}, {Anderson}, {Collier Cameron}, {Delrez}, {Gillon}, {Jehin}, {Lendl}, {Maxted}, {Pepe}, {Pollacco}, {Queloz}, {S{\'e}gransan}, {Smalley}, {Smith}, {Southworth}, {Triaud}, {Udry}, \& {West}}]{Hellier2014}
{Hellier}, C., {Anderson}, D.~R., {Collier Cameron}, A., {et~al.} 2014, MNRAS, 440, 1982, \dodoi{10.1093/mnras/stu410}

\bibitem[{{Hippke} {et~al.}(2019){Hippke}, {David}, {Mulders}, \& {Heller}}]{wotan2019}
{Hippke}, M., {David}, T.~J., {Mulders}, G.~D., \& {Heller}, R. 2019, AJ, 158, 143, \dodoi{10.3847/1538-3881/ab3984}

\bibitem[{{Hippke} \& {Heller}(2019)}]{tls2009}
{Hippke}, M., \& {Heller}, R. 2019, A\&A, 623, A39, \dodoi{10.1051/0004-6361/201834672}

\bibitem[{Ho \& Van Eylen(2023)}]{Ho2023}
Ho, C. S.~K., \& Van Eylen, V. 2023, Monthly Notices of the Royal Astronomical Society, 519, 4056, \dodoi{10.1093/mnras/stac3802}

\bibitem[{{Holmberg} {et~al.}(2009){Holmberg}, {Nordstr{\"o}m}, \& {Andersen}}]{Holmberg2009}
{Holmberg}, J., {Nordstr{\"o}m}, B., \& {Andersen}, J. 2009, A\&A, 501, 941, \dodoi{10.1051/0004-6361/200811191}

\bibitem[{{Howard} {et~al.}(2012){Howard}, {Marcy}, {Bryson}, {Jenkins}, {Rowe}, {Batalha}, {Borucki}, {Koch}, {Dunham}, {Gautier}, {Van Cleve}, {Cochran}, {Latham}, {Lissauer}, {Torres}, {Brown}, {Gilliland}, {Buchhave}, {Caldwell}, {Christensen-Dalsgaard}, {Ciardi}, {Fressin}, {Haas}, {Howell}, {Kjeldsen}, {Seager}, {Rogers}, {Sasselov}, {Steffen}, {Basri}, {Charbonneau}, {Christiansen}, {Clarke}, {Dupree}, {Fabrycky}, {Fischer}, {Ford}, {Fortney}, {Tarter}, {Girouard}, {Holman}, {Johnson}, {Klaus}, {Machalek}, {Moorhead}, {Morehead}, {Ragozzine}, {Tenenbaum}, {Twicken}, {Quinn}, {Isaacson}, {Shporer}, {Lucas}, {Walkowicz}, {Welsh}, {Boss}, {Devore}, {Gould}, {Smith}, {Morris}, {Prsa}, {Morton}, {Still}, {Thompson}, {Mullally}, {Endl}, \& {MacQueen}}]{Howard2012}
{Howard}, A.~W., {Marcy}, G.~W., {Bryson}, S.~T., {et~al.} 2012, ApJs, 201, 15, \dodoi{10.1088/0067-0049/201/2/15}

\bibitem[{{Hui-Bon-Hoa}(2021)}]{Hui2021}
{Hui-Bon-Hoa}, A. 2021, arXiv e-prints, arXiv:2101.08510, \dodoi{10.48550/arXiv.2101.08510}

\bibitem[{{Ida} \& {Lin}(2004)}]{Ida2004}
{Ida}, S., \& {Lin}, D.~N.~C. 2004, ApJ, 616, 567, \dodoi{10.1086/424830}

\bibitem[{{Johansen} {et~al.}(2007){Johansen}, {Oishi}, {Mac Low}, {Klahr}, {Henning}, \& {Youdin}}]{Johansen2007}
{Johansen}, A., {Oishi}, J.~S., {Mac Low}, M.-M., {et~al.} 2007, \nat, 448, 1022, \dodoi{10.1038/nature06086}

\bibitem[{{Johnson} {et~al.}(2010){Johnson}, {Aller}, {Howard}, \& {Crepp}}]{Johnson2010}
{Johnson}, J.~A., {Aller}, K.~M., {Howard}, A.~W., \& {Crepp}, J.~R. 2010, PASP, 122, 905, \dodoi{10.1086/655775}

\bibitem[{{Johnson} \& {Apps}(2009)}]{Johnson2009}
{Johnson}, J.~A., \& {Apps}, K. 2009, ApJ, 699, 933, \dodoi{10.1088/0004-637X/699/2/933}

\bibitem[{Johnson \& Li(2012)}]{Johnson2012}
Johnson, J.~L., \& Li, H. 2012, The Astrophysical Journal, 751, 81, \dodoi{10.1088/0004-637x/751/2/81}

\bibitem[{{Johnson} \& {Li}(2013)}]{Johnson2013}
{Johnson}, J.~L., \& {Li}, H. 2013, MNRAS, 431, 972, \dodoi{10.1093/mnras/stt229}

\bibitem[{{Jorge} {et~al.}(2022){Jorge}, {Kamp}, {Waters}, {Woitke}, \& {Spaargaren}}]{Jorge2022}
{Jorge}, D.~M., {Kamp}, I.~E.~E., {Waters}, L.~B.~F.~M., {Woitke}, P., \& {Spaargaren}, R.~J. 2022, A\&A, 660, A85, \dodoi{10.1051/0004-6361/202142738}

\bibitem[{{Kokubo} \& {Ida}(1998)}]{Kokubo1998}
{Kokubo}, E., \& {Ida}, S. 1998, \icarus, 131, 171, \dodoi{10.1006/icar.1997.5840}

\bibitem[{{Koppelman} {et~al.}(2018){Koppelman}, {Helmi}, \& {Veljanoski}}]{Koppelman2018}
{Koppelman}, H., {Helmi}, A., \& {Veljanoski}, J. 2018, ApJL, 860, L11, \dodoi{10.3847/2041-8213/aac882}

\bibitem[{{Kornet} {et~al.}(2005){Kornet}, {Bodenheimer}, {R{\'o}{\.z}yczka}, \& {Stepinski}}]{Kornet2005}
{Kornet}, K., {Bodenheimer}, P., {R{\'o}{\.z}yczka}, M., \& {Stepinski}, T.~F. 2005, A\&A, 430, 1133, \dodoi{10.1051/0004-6361:20041692}

\bibitem[{{Kov{\'a}cs} {et~al.}(2002){Kov{\'a}cs}, {Zucker}, \& {Mazeh}}]{Kovacs2002}
{Kov{\'a}cs}, G., {Zucker}, S., \& {Mazeh}, T. 2002, A\&A, 391, 369, \dodoi{10.1051/0004-6361:20020802}

\bibitem[{{Kratter} \& {Lodato}(2016)}]{Kratter2016}
{Kratter}, K., \& {Lodato}, G. 2016, \araa, 54, 271, \dodoi{10.1146/annurev-astro-081915-023307}

\bibitem[{{Kunimoto} \& {Bryson}(2020)}]{Kunimoto2020}
{Kunimoto}, M., \& {Bryson}, S. 2020, Research Notes of the American Astronomical Society, 4, 83, \dodoi{10.3847/2515-5172/ab9a3c}

\bibitem[{{Kutra} {et~al.}(2021){Kutra}, {Wu}, \& {Qian}}]{Kutra2021}
{Kutra}, T., {Wu}, Y., \& {Qian}, Y. 2021, AJ, 162, 69, \dodoi{10.3847/1538-3881/ac0431}

\bibitem[{{Lambrechts} \& {Johansen}(2014)}]{Lambrechts2014}
{Lambrechts}, M., \& {Johansen}, A. 2014, \aap, 572, A107, \dodoi{10.1051/0004-6361/201424343}

\bibitem[{{Lee}(2019)}]{Lee2019}
{Lee}, E.~J. 2019, \apj, 878, 36, \dodoi{10.3847/1538-4357/ab1b40}

\bibitem[{{Lee} \& {Chiang}(2015)}]{Lee2015}
{Lee}, E.~J., \& {Chiang}, E. 2015, \apj, 811, 41, \dodoi{10.1088/0004-637X/811/1/41}

\bibitem[{{Lee} {et~al.}(2014){Lee}, {Chiang}, \& {Ormel}}]{Lee2014}
{Lee}, E.~J., {Chiang}, E., \& {Ormel}, C.~W. 2014, \apj, 797, 95, \dodoi{10.1088/0004-637X/797/2/95}

\bibitem[{{Li} \& {Youdin}(2021)}]{Li2021}
{Li}, R., \& {Youdin}, A.~N. 2021, \apj, 919, 107, \dodoi{10.3847/1538-4357/ac0e9f}

\bibitem[{{Lin} {et~al.}(2018){Lin}, {Lee}, \& {Chiang}}]{Lin2018}
{Lin}, J.~W., {Lee}, E.~J., \& {Chiang}, E. 2018, \mnras, 480, 4338, \dodoi{10.1093/mnras/sty2159}

\bibitem[{{Lindegren} {et~al.}(2018){Lindegren}, {Hern{\'a}ndez}, {Bombrun}, {Klioner}, {Bastian}, {Ramos-Lerate}, {de Torres}, {Steidelm{\"u}ller}, {Stephenson}, {Hobbs}, {Lammers}, {Biermann}, {Geyer}, {Hilger}, {Michalik}, {Stampa}, {McMillan}, {Casta{\~n}eda}, {Clotet}, {Comoretto}, {Davidson}, {Fabricius}, {Gracia}, {Hambly}, {Hutton}, {Mora}, {Portell}, {van Leeuwen}, {Abbas}, {Abreu}, {Altmann}, {Andrei}, {Anglada}, {Balaguer-N{\'u}{\~n}ez}, {Barache}, {Becciani}, {Bertone}, {Bianchi}, {Bouquillon}, {Bourda}, {Br{\"u}semeister}, {Bucciarelli}, {Busonero}, {Buzzi}, {Cancelliere}, {Carlucci}, {Charlot}, {Cheek}, {Crosta}, {Crowley}, {de Bruijne}, {de Felice}, {Drimmel}, {Esquej}, {Fienga}, {Fraile}, {Gai}, {Garralda}, {Gonz{\'a}lez-Vidal}, {Guerra}, {Hauser}, {Hofmann}, {Holl}, {Jordan}, {Lattanzi}, {Lenhardt}, {Liao}, {Licata}, {Lister}, {L{\"o}ffler}, {Marchant}, {Martin-Fleitas}, {Messineo}, {Mignard}, {Morbidelli}, {Poggio}, {Riva}, {Rowell}, {Salguero}, {Sarasso}, {Sciacca}, {Siddiqui}, {Smart},
  {Spagna}, {Steele}, {Taris}, {Torra}, {van Elteren}, {van Reeven}, \& {Vecchiato}}]{Lindegren2018}
{Lindegren}, L., {Hern{\'a}ndez}, J., {Bombrun}, A., {et~al.} 2018, A\&A, 616, A2, \dodoi{10.1051/0004-6361/201832727}

\bibitem[{{Lodders}(2003)}]{Lodder2003}
{Lodders}, K. 2003, ApJ, 591, 1220, \dodoi{10.1086/375492}

\bibitem[{{Lodders}(2010)}]{Lodders2010}
{Lodders}, K. 2010, in Astrophysics and Space Science Proceedings, Vol.~16, Principles and Perspectives in Cosmochemistry, 379, \dodoi{10.1007/978-3-642-10352-0_8}

\bibitem[{{Lu} {et~al.}(2020){Lu}, {Schlaufman}, \& {Cheng}}]{Lu2020}
{Lu}, C.~X., {Schlaufman}, K.~C., \& {Cheng}, S. 2020, AJ, 160, 253, \dodoi{10.3847/1538-3881/abb773}

\bibitem[{{Luo} {et~al.}(2015){Luo}, {Zhao}, {Zhao}, {Deng}, {Liu}, {Jing}, {Wang}, {Zhang}, {Shi}, {Cui}, {Chu}, {Li}, {Bai}, {Wu}, {Cai}, {Cao}, {Cao}, {Carlin}, {Chen}, {Chen}, {Chen}, {Chen}, {Chen}, {Chen}, {Chen}, {Christlieb}, {Chu}, {Cui}, {Dong}, {Du}, {Fan}, {Feng}, {Fu}, {Gao}, {Gong}, {Gu}, {Guo}, {Han}, {He}, {Hou}, {Hou}, {Hou}, {Hu}, {Hu}, {Hu}, {Huo}, {Jia}, {Jiang}, {Jiang}, {Jiang}, {Jin}, {Kong}, {Kong}, {Lei}, {Li}, {Li}, {Li}, {Li}, {Li}, {Li}, {Li}, {Li}, {Li}, {Li}, {Li}, {Li}, {Liang}, {Lin}, {Liu}, {Liu}, {Liu}, {Liu}, {Lu}, {Luo}, {Mao}, {Newberg}, {Ni}, {Qi}, {Qi}, {Shen}, {Shi}, {Song}, {Song}, {Su}, {Su}, {Tang}, {Tao}, {Tian}, {Wang}, {Wang}, {Wang}, {Wang}, {Wang}, {Wang}, {Wang}, {Wang}, {Wang}, {Wang}, {Wang}, {Wang}, {Wang}, {Wang}, {Wang}, {Wang}, {Wang}, {Wang}, {Wang}, {Wang}, {Wei}, {Wei}, {Wu}, {Wu}, {Wu}, {Wu}, {Xing}, {Xu}, {Xu}, {Xu}, {Yan}, {Yang}, {Yang}, {Yang}, {Yang}, {Yao}, {Yu}, {Yuan}, {Yuan}, {Yuan}, {Yuan}, {Zhai}, {Zhang}, {Zhang}, {Zhang}, {Zhang},
  {Zhang}, {Zhang}, {Zhang}, {Zhang}, {Zhao}, {Zhou}, {Zhou}, {Zhu}, {Zhu}, {Zou}, \& {Zuo}}]{Luo2015}
{Luo}, A.~L., {Zhao}, Y.-H., {Zhao}, G., {et~al.} 2015, Research in Astronomy and Astrophysics, 15, 1095, \dodoi{10.1088/1674-4527/15/8/002}

\bibitem[{{Marcy} {et~al.}(2005){Marcy}, {Butler}, {Fischer}, {Vogt}, {Wright}, {Tinney}, \& {Jones}}]{Marcy2005}
{Marcy}, G., {Butler}, R.~P., {Fischer}, D., {et~al.} 2005, Progress of Theoretical Physics Supplement, 158, 24, \dodoi{10.1143/PTPS.158.24}

\bibitem[{{Masuda} \& {Winn}(2017)}]{Masuda2017}
{Masuda}, K., \& {Winn}, J.~N. 2017, AJ, 153, 187, \dodoi{10.3847/1538-3881/aa647c}

\bibitem[{{Mayor} {et~al.}(2011){Mayor}, {Marmier}, {Lovis}, {Udry}, {S{\'e}gransan}, {Pepe}, {Benz}, {Bertaux}, {Bouchy}, {Dumusque}, {Lo Curto}, {Mordasini}, {Queloz}, \& {Santos}}]{Mayor2011}
{Mayor}, M., {Marmier}, M., {Lovis}, C., {et~al.} 2011, arXiv e-prints, arXiv:1109.2497.
\newblock \doarXiv{1109.2497}

\bibitem[{{McCarthy} \& {Zuckerman}(2004)}]{McCarthy2004}
{McCarthy}, C., \& {Zuckerman}, B. 2004, AJ, 127, 2871, \dodoi{10.1086/383559}

\bibitem[{{Ment} \& {Charbonneau}(2023)}]{Ment2023}
{Ment}, K., \& {Charbonneau}, D. 2023, \aj, 165, 265, \dodoi{10.3847/1538-3881/acd175}

\bibitem[{{Miller} \& {Fortney}(2011)}]{Miller2011}
{Miller}, N., \& {Fortney}, J.~J. 2011, ApJL, 736, L29, \dodoi{10.1088/2041-8205/736/2/L29}

\bibitem[{{Moe} \& {Kratter}(2019)}]{Maxwell2019}
{Moe}, M., \& {Kratter}, K.~M. 2019, arXiv e-prints, arXiv:1912.01699.
\newblock \doarXiv{1912.01699}

\bibitem[{{Moe} {et~al.}(2019){Moe}, {Kratter}, \& {Badenes}}]{Moe2019}
{Moe}, M., {Kratter}, K.~M., \& {Badenes}, C. 2019, \apj, 875, 61, \dodoi{10.3847/1538-4357/ab0d88}

\bibitem[{{Mordasini} {et~al.}(2012){Mordasini}, {Alibert}, {Benz}, {Klahr}, \& {Henning}}]{Mordasini2012}
{Mordasini}, C., {Alibert}, Y., {Benz}, W., {Klahr}, H., \& {Henning}, T. 2012, A\&A, 541, A97, \dodoi{10.1051/0004-6361/201117350}

\bibitem[{{Mortier} {et~al.}(2013){Mortier}, {Santos}, {Sousa}, {Israelian}, {Mayor}, \& {Udry}}]{mortier13}
{Mortier}, A., {Santos}, N.~C., {Sousa}, S., {et~al.} 2013, A\&A, 551, A112, \dodoi{10.1051/0004-6361/201220707}

\bibitem[{{Mortier} {et~al.}(2012){Mortier}, {Santos}, {Sozzetti}, {Mayor}, {Latham}, {Bonfils}, \& {Udry}}]{mortier2012}
{Mortier}, A., {Santos}, N.~C., {Sozzetti}, A., {et~al.} 2012, A\&A, 543, A45, \dodoi{10.1051/0004-6361/201118651}

\bibitem[{{Mulders} {et~al.}(2015){Mulders}, {Pascucci}, \& {Apai}}]{Mulders2015}
{Mulders}, G.~D., {Pascucci}, I., \& {Apai}, D. 2015, ApJ, 798, 112, \dodoi{10.1088/0004-637X/798/2/112}

\bibitem[{{Mulders} {et~al.}(2018){Mulders}, {Pascucci}, {Apai}, \& {Ciesla}}]{Mulders2018}
{Mulders}, G.~D., {Pascucci}, I., {Apai}, D., \& {Ciesla}, F.~J. 2018, AJ, 156, 24, \dodoi{10.3847/1538-3881/aac5ea}

\bibitem[{{Omukai}(2000)}]{Omukai2000}
{Omukai}, K. 2000, ApJ, 534, 809, \dodoi{10.1086/308776}

\bibitem[{{Ormel} \& {Klahr}(2010)}]{Ormel2010}
{Ormel}, C.~W., \& {Klahr}, H.~H. 2010, \aap, 520, A43, \dodoi{10.1051/0004-6361/201014903}

\bibitem[{{Petigura} {et~al.}(2018){Petigura}, {Marcy}, {Winn}, {Weiss}, {Fulton}, {Howard}, {Sinukoff}, {Isaacson}, {Morton}, \& {Johnson}}]{Petigura2018}
{Petigura}, E.~A., {Marcy}, G.~W., {Winn}, J.~N., {et~al.} 2018, AJ, 155, 89, \dodoi{10.3847/1538-3881/aaa54c}

\bibitem[{Pettitt(1976)}]{Pettitt1976}
Pettitt, A.~N. 1976, Biometrika, 63, 161.
\newblock \url{http://www.jstor.org/stable/2335097}

\bibitem[{{Piso} \& {Youdin}(2014)}]{Piso2014}
{Piso}, A.-M.~A., \& {Youdin}, A.~N. 2014, \apj, 786, 21, \dodoi{10.1088/0004-637X/786/1/21}

\bibitem[{{Pollack} {et~al.}(1996){Pollack}, {Hubickyj}, {Bodenheimer}, {Lissauer}, {Podolak}, \& {Greenzweig}}]{Pollack1996}
{Pollack}, J.~B., {Hubickyj}, O., {Bodenheimer}, P., {et~al.} 1996, \icarus, 124, 62, \dodoi{10.1006/icar.1996.0190}

\bibitem[{{Pont} {et~al.}(2006){Pont}, {Zucker}, \& {Queloz}}]{Pont2006}
{Pont}, F., {Zucker}, S., \& {Queloz}, D. 2006, MNRAS, 373, 231, \dodoi{10.1111/j.1365-2966.2006.11012.x}

\bibitem[{{Rafikov}(2006)}]{Rafikov2006}
{Rafikov}, R.~R. 2006, \apj, 648, 666, \dodoi{10.1086/505695}

\bibitem[{{Ricker} {et~al.}(2015{\natexlab{a}}){Ricker}, {Winn}, {Vanderspek}, {Latham}, {Bakos}, {Bean}, {Berta-Thompson}, {Brown}, {Buchhave}, {Butler}, {Butler}, {Chaplin}, {Charbonneau}, {Christensen-Dalsgaard}, {Clampin}, {Deming}, {Doty}, {De Lee}, {Dressing}, {Dunham}, {Endl}, {Fressin}, {Ge}, {Henning}, {Holman}, {Howard}, {Ida}, {Jenkins}, {Jernigan}, {Johnson}, {Kaltenegger}, {Kawai}, {Kjeldsen}, {Laughlin}, {Levine}, {Lin}, {Lissauer}, {MacQueen}, {Marcy}, {McCullough}, {Morton}, {Narita}, {Paegert}, {Palle}, {Pepe}, {Pepper}, {Quirrenbach}, {Rinehart}, {Sasselov}, {Sato}, {Seager}, {Sozzetti}, {Stassun}, {Sullivan}, {Szentgyorgyi}, {Torres}, {Udry}, \& {Villasenor}}]{TESS}
{Ricker}, G.~R., {Winn}, J.~N., {Vanderspek}, R., {et~al.} 2015{\natexlab{a}}, Journal of Astronomical Telescopes, Instruments, and Systems, 1, 014003, \dodoi{10.1117/1.JATIS.1.1.014003}

\bibitem[{{Ricker} {et~al.}(2015{\natexlab{b}}){Ricker}, {Winn}, {Vanderspek}, {Latham}, {Bakos}, {Bean}, {Berta-Thompson}, {Brown}, {Buchhave}, {Butler}, {Butler}, {Chaplin}, {Charbonneau}, {Christensen-Dalsgaard}, {Clampin}, {Deming}, {Doty}, {De Lee}, {Dressing}, {Dunham}, {Endl}, {Fressin}, {Ge}, {Henning}, {Holman}, {Howard}, {Ida}, {Jenkins}, {Jernigan}, {Johnson}, {Kaltenegger}, {Kawai}, {Kjeldsen}, {Laughlin}, {Levine}, {Lin}, {Lissauer}, {MacQueen}, {Marcy}, {McCullough}, {Morton}, {Narita}, {Paegert}, {Palle}, {Pepe}, {Pepper}, {Quirrenbach}, {Rinehart}, {Sasselov}, {Sato}, {Seager}, {Sozzetti}, {Stassun}, {Sullivan}, {Szentgyorgyi}, {Torres}, {Udry}, \& {Villasenor}}]{Ricker2015}
---. 2015{\natexlab{b}}, Journal of Astronomical Telescopes, Instruments, and Systems, 1, 014003, \dodoi{10.1117/1.JATIS.1.1.014003}

\bibitem[{{Rosotti} {et~al.}(2020){Rosotti}, {Teague}, {Dullemond}, {Booth}, \& {Clarke}}]{Rosotti20}
{Rosotti}, G.~P., {Teague}, R., {Dullemond}, C., {Booth}, R.~A., \& {Clarke}, C.~J. 2020, \mnras, 495, 173, \dodoi{10.1093/mnras/staa1170}

\bibitem[{{Sandford} \& {Kipping}(2017)}]{Sandford2017}
{Sandford}, E., \& {Kipping}, D. 2017, \aj, 154, 228, \dodoi{10.3847/1538-3881/aa94bf}

\bibitem[{{Santerne} {et~al.}(2016){Santerne}, {Moutou}, {Tsantaki}, {Bouchy}, {H{\'e}brard}, {Adibekyan}, {Almenara}, {Amard}, {Barros}, {Boisse}, {Bonomo}, {Bruno}, {Courcol}, {Deleuil}, {Demangeon}, {D{\'\i}az}, {Guillot}, {Havel}, {Montagnier}, {Rajpurohit}, {Rey}, \& {Santos}}]{Santerne}
{Santerne}, A., {Moutou}, C., {Tsantaki}, M., {et~al.} 2016, A\&A, 587, A64, \dodoi{10.1051/0004-6361/201527329}

\bibitem[{{Santos} {et~al.}(2004){Santos}, {Israelian}, \& {Mayor}}]{Santos2004}
{Santos}, N.~C., {Israelian}, G., \& {Mayor}, M. 2004, A\&A, 415, 1153, \dodoi{10.1051/0004-6361:20034469}

\bibitem[{{Schneider} {et~al.}(2002){Schneider}, {Ferrara}, {Natarajan}, \& {Omukai}}]{Schneider2002}
{Schneider}, R., {Ferrara}, A., {Natarajan}, P., \& {Omukai}, K. 2002, ApJ, 571, 30, \dodoi{10.1086/339917}

\bibitem[{{Seager} \& {Mall{\'e}n-Ornelas}(2003)}]{Seager2003}
{Seager}, S., \& {Mall{\'e}n-Ornelas}, G. 2003, \apj, 585, 1038, \dodoi{10.1086/346105}

\bibitem[{{Sharma} {et~al.}(2011){Sharma}, {Bland-Hawthorn}, {Johnston}, \& {Binney}}]{Sharma2011}
{Sharma}, S., {Bland-Hawthorn}, J., {Johnston}, K.~V., \& {Binney}, J. 2011, ApJ, 730, 3, \dodoi{10.1088/0004-637X/730/1/3}

\bibitem[{{Sharma} {et~al.}(2018){Sharma}, {Stello}, {Buder}, {Kos}, {Bland-Hawthorn}, {Asplund}, {Duong}, {Lin}, {Lind}, {Ness}, {Huber}, {Zwitter}, {Traven}, {Hon}, {Kafle}, {Khanna}, {Saddon}, {Anguiano}, {Casey}, {Freeman}, {Martell}, {De Silva}, {Simpson}, {Wittenmyer}, \& {Zucker}}]{Sharma2018}
{Sharma}, S., {Stello}, D., {Buder}, S., {et~al.} 2018, MNRAS, 473, 2004, \dodoi{10.1093/mnras/stx2582}

\bibitem[{{Soubiran} {et~al.}(2008){Soubiran}, {Bienaym{\'e}}, {Mishenina}, \& {Kovtyukh}}]{Soubiran2008}
{Soubiran}, C., {Bienaym{\'e}}, O., {Mishenina}, T.~V., \& {Kovtyukh}, V.~V. 2008, \aap, 480, 91, \dodoi{10.1051/0004-6361:20078788}

\bibitem[{{Soubiran} {et~al.}(2022){Soubiran}, {Brouillet}, \& {Casamiquela}}]{Soubiran2022}
{Soubiran}, C., {Brouillet}, N., \& {Casamiquela}, L. 2022, A\&A, 663, A4, \dodoi{10.1051/0004-6361/202142409}

\bibitem[{{Soubiran} {et~al.}(2010){Soubiran}, {Le Campion}, {Cayrel de Strobel}, \& {Caillo}}]{Soubiran2010}
{Soubiran}, C., {Le Campion}, J.~F., {Cayrel de Strobel}, G., \& {Caillo}, A. 2010, A\&A, 515, A111, \dodoi{10.1051/0004-6361/201014247}

\bibitem[{{Sozzetti} {et~al.}(2009){Sozzetti}, {Torres}, {Latham}, {Stefanik}, {Korzennik}, {Boss}, {Carney}, \& {Laird}}]{sozzetti2009}
{Sozzetti}, A., {Torres}, G., {Latham}, D.~W., {et~al.} 2009, ApJ, 697, 544, \dodoi{10.1088/0004-637X/697/1/544}

\bibitem[{{Squire} \& {Hopkins}(2018)}]{Squire2018}
{Squire}, J., \& {Hopkins}, P.~F. 2018, \mnras, 477, 5011, \dodoi{10.1093/mnras/sty854}

\bibitem[{{Stassun} \& {Torres}(2021)}]{Stassun2021}
{Stassun}, K.~G., \& {Torres}, G. 2021, ApJL, 907, L33, \dodoi{10.3847/2041-8213/abdaad}

\bibitem[{{Stassun} {et~al.}(2018{\natexlab{a}}){Stassun}, {Oelkers}, {Pepper}, {Paegert}, {De Lee}, {Torres}, {Latham}, {Charpinet}, {Dressing}, {Huber}, {Kane}, {L{\'e}pine}, {Mann}, {Muirhead}, {Rojas-Ayala}, {Silvotti}, {Fleming}, {Levine}, \& {Plavchan}}]{Stassun2018}
{Stassun}, K.~G., {Oelkers}, R.~J., {Pepper}, J., {et~al.} 2018{\natexlab{a}}, AJ, 156, 102, \dodoi{10.3847/1538-3881/aad050}

\bibitem[{{Stassun} {et~al.}(2018{\natexlab{b}}){Stassun}, {Oelkers}, {Pepper}, {Paegert}, {De Lee}, {Torres}, {Latham}, {Charpinet}, {Dressing}, {Huber}, {Kane}, {L{\'e}pine}, {Mann}, {Muirhead}, {Rojas-Ayala}, {Silvotti}, {Fleming}, {Levine}, \& {Plavchan}}]{Statsun2018}
---. 2018{\natexlab{b}}, AJ, 156, 102, \dodoi{10.3847/1538-3881/aad050}

\bibitem[{{Stevenson}(1982)}]{Stevenson1982}
{Stevenson}, D.~J. 1982, \planss, 30, 755, \dodoi{10.1016/0032-0633(82)90108-8}

\bibitem[{{Thorngren} {et~al.}(2019){Thorngren}, {Marley}, \& {Fortney}}]{Thorngren2019}
{Thorngren}, D.~P., {Marley}, M.~S., \& {Fortney}, J.~J. 2019, Research Notes of the American Astronomical Society, 3, 128, \dodoi{10.3847/2515-5172/ab4353}

\bibitem[{{Thuillier} {et~al.}(2022){Thuillier}, {Van Grootel}, {D{\'e}vora-Pajares}, {Pozuelos}, {Charpinet}, \& {Siess}}]{Thuillier2022}
{Thuillier}, A., {Van Grootel}, V., {D{\'e}vora-Pajares}, M., {et~al.} 2022, \aap, 664, A113, \dodoi{10.1051/0004-6361/202243554}

\bibitem[{{Tokovinin} {et~al.}(2013){Tokovinin}, {Fischer}, {Bonati}, {Giguere}, {Moore}, {Schwab}, {Spronck}, \& {Szymkowiak}}]{Tokovinin2013}
{Tokovinin}, A., {Fischer}, D.~A., {Bonati}, M., {et~al.} 2013, PASP, 125, 1336, \dodoi{10.1086/674012}

\bibitem[{{Tsiaras} {et~al.}(2016){Tsiaras}, {Waldmann}, {Rocchetto}, {Varley}, {Morello}, {Damiano}, \& {Tinetti}}]{pylight}
{Tsiaras}, A., {Waldmann}, I.~P., {Rocchetto}, M., {et~al.} 2016, {pylightcurve: Exoplanet lightcurve model}.
\newblock \doeprint{1612.018}

\bibitem[{{Udry} \& {Santos}(2007)}]{Udry2007}
{Udry}, S., \& {Santos}, N.~C. 2007, ARA\&A, 45, 397, \dodoi{10.1146/annurev.astro.45.051806.110529}

\bibitem[{{Van Eylen} {et~al.}(2018){Van Eylen}, {Agentoft}, {Lundkvist}, {Kjeldsen}, {Owen}, {Fulton}, {Petigura}, \& {Snellen}}]{VanEylen2018}
{Van Eylen}, V., {Agentoft}, C., {Lundkvist}, M.~S., {et~al.} 2018, MNRAS, 479, 4786, \dodoi{10.1093/mnras/sty1783}

\bibitem[{{Venturini} {et~al.}(2020){Venturini}, {Guilera}, {Ronco}, \& {Mordasini}}]{Venturini2020}
{Venturini}, J., {Guilera}, O.~M., {Ronco}, M.~P., \& {Mordasini}, C. 2020, A\&A, 644, A174, \dodoi{10.1051/0004-6361/202039140}

\bibitem[{Vugrin {et~al.}(2007)Vugrin, Swiler, Roberts, Stucky-Mack, \& Sullivan}]{Vugrin2007}
Vugrin, K.~W., Swiler, L.~P., Roberts, R.~M., Stucky-Mack, N.~J., \& Sullivan, S.~P. 2007, Water Resources Research, 43, \dodoi{https://doi.org/10.1029/2005WR004804}

\bibitem[{{Wang} {et~al.}(2015){Wang}, {Fischer}, {Horch}, \& {Huang}}]{Wang2015}
{Wang}, J., {Fischer}, D.~A., {Horch}, E.~P., \& {Huang}, X. 2015, ApJ, 799, 229, \dodoi{10.1088/0004-637X/799/2/229}

\bibitem[{{Wright} {et~al.}(2012){Wright}, {Marcy}, {Howard}, {Johnson}, {Morton}, \& {Fischer}}]{Wright2012}
{Wright}, J.~T., {Marcy}, G.~W., {Howard}, A.~W., {et~al.} 2012, ApJ, 753, 160, \dodoi{10.1088/0004-637X/753/2/160}

\bibitem[{{Wyatt} {et~al.}(2007){Wyatt}, {Clarke}, \& {Greaves}}]{Wyatt2007}
{Wyatt}, M.~C., {Clarke}, C.~J., \& {Greaves}, J.~S. 2007, MNRAS, 380, 1737, \dodoi{10.1111/j.1365-2966.2007.12244.x}

\bibitem[{{Xin} {et~al.}(2022){Xin}, {Renzo}, \& {Metzger}}]{Xin2022}
{Xin}, C., {Renzo}, M., \& {Metzger}, B.~D. 2022, \mnras, 516, 5816, \dodoi{10.1093/mnras/stac2551}

\bibitem[{{Yasui} {et~al.}(2010){Yasui}, {Kobayashi}, {Tokunaga}, {Saito}, \& {Tokoku}}]{Yasui2010}
{Yasui}, C., {Kobayashi}, N., {Tokunaga}, A.~T., {Saito}, M., \& {Tokoku}, C. 2010, ApJL, 723, L113, \dodoi{10.1088/2041-8205/723/1/L113}

\bibitem[{{Youdin} \& {Goodman}(2005)}]{Youdin2005}
{Youdin}, A.~N., \& {Goodman}, J. 2005, \apj, 620, 459, \dodoi{10.1086/426895}

\bibitem[{{Zechmeister} \& {K{\"u}rster}(2009)}]{zechmeister2009}
{Zechmeister}, M., \& {K{\"u}rster}, M. 2009, A\&A, 496, 577, \dodoi{10.1051/0004-6361:200811296}

\bibitem[{{Zhou} {et~al.}(2017){Zhou}, {Bakos}, {Hartman}, {Latham}, {Torres}, {Bhatti}, {Penev}, {Buchhave}, {Kov{\'a}cs}, {Bieryla}, {Quinn}, {Isaacson}, {Fulton}, {Falco}, {Csubry}, {Everett}, {Szklenar}, {Esquerdo}, {Berlind}, {Calkins}, {B{\'e}ky}, {Knox}, {Hinz}, {Horch}, {Hirsch}, {Howell}, {Noyes}, {Marcy}, {de Val-Borro}, {L{\'a}z{\'a}r}, {Papp}, \& {S{\'a}ri}}]{Zhou2017}
{Zhou}, G., {Bakos}, G.~{\'A}., {Hartman}, J.~D., {et~al.} 2017, AJ, 153, 211, \dodoi{10.3847/1538-3881/aa674a}

\bibitem[{{Zhou} {et~al.}(2019){Zhou}, {Huang}, {Bakos}, {Hartman}, {Latham}, {Quinn}, {Collins}, {Winn}, {Wong}, {Kov{\'a}cs}, {Csubry}, {Bhatti}, {Penev}, {Bieryla}, {Esquerdo}, {Berlind}, {Calkins}, {de Val-Borro}, {Noyes}, {L{\'a}z{\'a}r}, {Papp}, {S{\'a}ri}, {Kov{\'a}cs}, {Buchhave}, {Szklenar}, {B{\'e}ky}, {Johnson}, {Cochran}, {Kniazev}, {Stassun}, {Fulton}, {Shporer}, {Espinoza}, {Bayliss}, {Everett}, {Howell}, {Hellier}, {Anderson}, {Collier Cameron}, {West}, {Brown}, {Schanche}, {Barkaoui}, {Pozuelos}, {Gillon}, {Jehin}, {Benkhaldoun}, {Daassou}, {Ricker}, {Vanderspek}, {Seager}, {Jenkins}, {Lissauer}, {Armstrong}, {Collins}, {Gan}, {Hart}, {Horne}, {Kielkopf}, {Nielsen}, {Nishiumi}, {Narita}, {Palle}, {Relles}, {Sefako}, {Tan}, {Davies}, {Goeke}, {Guerrero}, {Haworth}, \& {Villanueva}}]{Zhou2020}
{Zhou}, G., {Huang}, C.~X., {Bakos}, G.~{\'A}., {et~al.} 2019, AJ, 158, 141, \dodoi{10.3847/1538-3881/ab36b5}

\bibitem[{{Zhu}(2019)}]{Zhu2019}
{Zhu}, W. 2019, ApJ, 873, 8, \dodoi{10.3847/1538-4357/ab0205}

\bibitem[{{Zink} {et~al.}(2019){Zink}, {Christiansen}, \& {Hansen}}]{ExoMult}
{Zink}, J.~K., {Christiansen}, J.~L., \& {Hansen}, B. M.~S. 2019, MNRAS, 483, 4479, \dodoi{10.1093/mnras/sty3463}

\bibitem[{{Zink} {et~al.}(2020{\natexlab{a}}){Zink}, {Hardegree-Ullman}, {Christiansen}, {Dressing}, {Crossfield}, {Petigura}, {Schlieder}, \& {Ciardi}}]{edi}
{Zink}, J.~K., {Hardegree-Ullman}, K.~K., {Christiansen}, J.~L., {et~al.} 2020{\natexlab{a}}, AJ, 159, 154, \dodoi{10.3847/1538-3881/ab7448}

\bibitem[{{Zink} {et~al.}(2020{\natexlab{b}}){Zink}, {Hardegree-Ullman}, {Christiansen}, {Petigura}, {Dressing}, {Schlieder}, {Ciardi}, \& {Crossfield}}]{Zink2020A}
---. 2020{\natexlab{b}}, AJ, 160, 94, \dodoi{10.3847/1538-3881/aba123}

\bibitem[{{Zink} {et~al.}(2023){Zink}, {Hardegree-Ullman}, {Christiansen}, {Petigura}, {Boley}, {Bhure}, {Rice}, {Yee}, {Isaacson}, {Fernandes}, {Howard}, {Blunt}, {Lubin}, {Chontos}, {Pidhorodetska}, \& {MacDougall}}]{Zink2023}
---. 2023, AJ, 165, 262, \dodoi{10.3847/1538-3881/acd24c}

\end{thebibliography}
\nocite{*}

\end{CJK*}
\end{document}